 \documentclass[final,3p]{elsarticle}

\usepackage{amssymb}
\usepackage{amsmath}

\usepackage{siunitx} 
\DeclareSIUnit{\ab}{\text{ab}^{-1}}

\usepackage{subcaption}

\usepackage[hidelinks]{hyperref}

\journal{Nuclear Inst. and Methods in Physics Research, A}

\begin{document}

\begin{frontmatter}



\title{Monochromatization interaction region optics design for direct \textit{s}-channel Higgs production at FCC-ee} 

 \author[label1,label2,label3]{Z. Zhang\corref{cor1}}
 \cortext[cor1]{Corresponding author}
 \ead{zhangzhandong@ihep.ac.cn, zhandong.zhang@ijclab.in2p3.fr}
 
 \author[label2]{A. Faus-Golfe}
 \author[label2]{A. Korsun}
 \author[label4]{B. Bai}
 \author[label5]{H. Jiang}
 \author[label6,label7]{K. Oide}
 \author[label8]{P. Raimondi}
 \author[label9]{D. d’Enterria}
 \author[label1]{S. Zhang}
 \author[label1,label3]{Z. Zhou}
 \author[label1]{Y. Chi}
 \author[label9]{F. Zimmermann}
 
 \affiliation[label1]{organization={Institute of High Energy Physics, Chinese Academy of Sciences},
             city={Beijing 100049},
             country={China}}

 \affiliation[label2]{organization={Laboratoire de Physique des 2 Infinis Irène Joliot-Curie (IJCLab), CNRS/IN2P3,\\Université Paris-Saclay},
             city={91405 Orsay},
             country={France}}

 \affiliation[label3]{organization={University of Chinese Academy of Sciences},
             city={Beijing 100049},
             country={China}}

 \affiliation[label4]{organization={Institute of Special Environments Physical Sciences, Harbin Institute of Technology (Shenzhen)},
             city={Shenzhen 518055},
             country={China}}

 \affiliation[label5]{organization={Lancaster University},
             city={Lancaster LA1 4YW},
             country={UK}}

 \affiliation[label6]{organization={Université de Genève},
             city={1211 Geneva 4},
             country={Switzerland}}

 \affiliation[label7]{organization={High Energy Accelerator Research Organization (KEK) 1-1 Oho},
             city={Tsukuba},
             postcode={Ibaraki 305-0801},
             country={Japan}}

 \affiliation[label8]{organization={Fermi National Accelerator Laboratory},
             city={Batavia},
             postcode={IL 60510},
             country={USA}}

 \affiliation[label9]{organization={European Organization for Nuclear Research (CERN)},
             city={1211 Geneva 23},
             country={Switzerland}}



\begin{abstract}
The FCC-ee offers the potential to measure the electron Yukawa coupling via direct \textit{s}-channel Higgs production, $e^+ e^- \rightarrow \text{H}$, at a centre-of-mass (CM) energy of $\sim$\SI{125}{\giga\electronvolt}. This measurement is significantly facilitated if the CM energy spread of $e^+ e^-$ collisions can be reduced to a level comparable to the natural width of the Higgs boson, $\Gamma_{\text{H}}=\SI{4.1}{\mega\electronvolt}$, without substantial loss in luminosity. Achieving this reduction in collision-energy spread is possible through the “monochromatization” concept. The basic idea is to create opposite correlations between spatial position and energy deviation within the colliding beams, which can be accomplished in beam optics by introducing a nonzero dispersion function with opposite signs for the two beams at the interaction point. Since the first proposal in 2016, the implementation of monochromatization at the FCC-ee has been continuously improved, starting from preliminary parametric studies. In this paper, we present a detailed study of the interaction region optics design for this newly proposed collision mode, exploring different potential configurations and their implementation in the FCC-ee global lattice, along with beam dynamics simulations and performance evaluations including the impact of “beamstrahlung.”
\end{abstract}



\begin{keyword}
FCC-ee \sep monochromatization \sep optics design \sep luminosity \sep centre-of-mass energy spread \sep beamstrahlung


\end{keyword}

\end{frontmatter}



\section{Introduction}
\label{sec1}

One of the most fundamental outstanding measurements since the Higgs boson discovery~\cite{ATLAS:2012yve,CMS:2012qbp} is determining its Yukawa couplings~\cite{FCC:2018byv,Abada2019FCCeeTL,Jadach:2015cwa}. Measuring the coupling of first-generation fermions (u and d quarks, and electron) presents significant experimental challenges due to their low masses and consequently small Yukawa couplings to the Higgs field. The Yukawa coupling of the electron, $y_{e} = \sqrt{2} m_{e}/v = \SI{2.9e-6}{}$ (with the electron rest mass energy $m_{e} = \SI{0.511e-3}{\giga\electronvolt}$ and Higgs field vacuum expectation value $v=(\sqrt{2}G_{F})^{-1/2}=\SI{246.22}{\giga\electronvolt}$), is virtually impossible to measure at hadron colliders because the $\text{H} \rightarrow e^+ e^-$ decay has a tiny branching fraction of $\mathcal{B}(\text{H} \rightarrow e^+ e^-) = \SI{5.22e-9}{}$ and it is completely swamped by the Drell-Yan dielectron continuum with many orders of magnitude larger cross section. Around ten years ago, it was first noticed that an unparalleled integrated luminosity ($\mathcal{L}_{\text{int}}$) of \SI{10}{\ab} per year, achievable at the future high-energy circular $e^+ e^-$ collider FCC-ee~\cite{FCC:2018byv,Abada2019FCCeeTL} at a centre-of-mass (CM) energy of $\sim$\SI{125}{\giga\electronvolt}, could enable an attempt to observe the resonant \textit{s}-channel production of the scalar boson, namely the reaction $e^+ e^- \rightarrow \text{H}$ at the Higgs pole~\cite{dEnterria:2014,dEnterria:2017dac,dEnterria:2021xij}. This motivation led to subsequent theoretical~\cite{Jadach:2015cwa,Altmannshofer2015ExperimentalCO} and accelerator~\cite{ValdiviaGarcia:2016rdg,ValdiviaGarcia:2016fae,ValdiviaGarcia:2017xiv,Bogomyagkov:2017tpk,Shatilov:2018drj,Zimmermann:2017tjv,ValdiviaGarcia:2018cpm,ValdiviaGarcia:2019ezi,telnov2020monochromatizationeecolliderslarge,Garcia:2019nci,ValdiviaGarcia:2022nks,Faus-Golfe:2021udx} studies of the $e^+ e^- \rightarrow \text{H}$ process.

\begin{figure}[h!]
\centering
\includegraphics[width=0.5\textwidth]{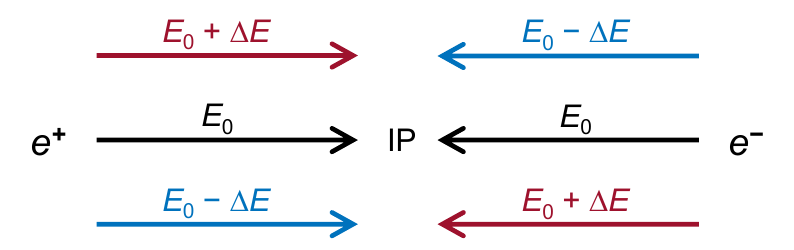}
\caption{Head-on collision scheme with monochromatization.}\label{fig1}
\end{figure}

Such a measurement is more easily feasible if the CM energy spread ($\sigma_{W}$) of  $e^+ e^-$ collisions can be reduced to a level comparable to the natural width of the Standard Model (SM) Higgs boson, $\Gamma_{\text{H}}=\SI{4.1}{\mega\electronvolt}$. However, in a conventional collision scheme, the natural collision-energy spread at \SI{125}{\giga\electronvolt}, caused solely by synchrotron radiation (SR), is $\sim$\SI{50}{\mega\electronvolt}. Reducing it to the desired level, thus enhancing the energy resolution in colliding-beam experiments, could be achieved by exploiting the concept of monochromatization~\cite{Renieri:1975wt}. The basic idea consists of generating opposite correlations between spatial position and energy deviation within the colliding beams, as illustrated in Fig.~\ref{fig1} with the nominal beam energy ($E_0$).

In this configuration, the CM energy $W = 2E_0 + \mathcal{O}(\Delta E)^2$ is reduced without necessarily decreasing the inherent energy spread of the two individual beams. In terms of beam optics, this can be achieved by generating a nonzero dispersion function with opposite signs for the two beams at the interaction point (IP). This requires adding the necessary components within the interaction region (IR) of the collider to create such a dispersion function. A nonzero dispersion function at the IP ($D_{x,y}^{\ast} \neq 0$) enlarges the IP transverse beam size ($\sigma_{x,y}^{\ast}$) which in turn affects the luminosity ($\mathcal{L} \propto 1/(\sigma_{x}^{\ast} \sigma_{y}^{\ast})$). A monochromatization factor ($\lambda$) can be introduced as:
\begin{equation}
    \lambda = \sqrt{1 + \sigma_{\delta}^2 \left( \frac{D_x^{\ast 2}}{\varepsilon_{x} \beta_{x}^{\ast}} + \frac{D_y^{\ast 2}}{\varepsilon_{y} \beta_{y}^{\ast}} \right) }
\label{eq1}
\end{equation}
with $\sigma_{\delta}$ the relative energy spread, $\varepsilon_{x,y}$ the transverse emittances and $\beta_{x,y}^{\ast}$ the betatron functions at the IP. Using $\lambda$, the $\sigma_{W}$ and the $\mathcal{L}$ in the monochromatization operation mode are given by:
\begin{equation}
    \sigma_W = \frac{\sqrt{2} E_0 \sigma_\delta}{\lambda} \: \text{and} \: \mathcal{L} = \frac{\mathcal{L}_0}{\lambda}\:,
\label{eq2}
\end{equation}
where $\mathcal{L}_0$ represents the luminosity for the same values of $\beta_{x,y}^{\ast}$ but without $D_{x,y}^{\ast}$. Consequently, the design of a monochromatization scheme requires consideration of both the IR beam optics and the optimization of other collider parameters to maintain the highest possible luminosity.






The essential concept and the basic theory of monochromatization in $e^+ e^-$ colliders were first proposed in 1975 by A. Renieri, with the aim of improving the energy resolution of ADONE~\cite{Renieri:1975wt,Bassetti:1974if}. Since then, a variety of monochromatization schemes using different techniques for the generation of the dispersion function in the IP, including electrostatic quadrupoles, Radio Frequency (RF) magnetic devices, dipole magnets with electrostatic separators or superconducting RF deflecting (SCRF-D) cavities, have been proposed for multiple circular $e^+ e^-$ colliders, mostly at low energies and in head-on collision configuration: VEPP-4~\cite{Protopopov:1979tn,Avdienko:1983mee}, SPEAR~\cite{Wille:1984fb}, LEP~\cite{Jowett:1985zx,Bassetti:1987zg,Jowett:806265}, B factory~\cite{DUBROVIN:1991,Veshcherevich:1992by} or Tau-Charm factories~\cite{Alexahin:1990zps,Zholents:1992ua,Faus-Golfe:1996uai,Schulte:1997nga,Zholents:1988bu}. Despite its theoretical simplicity, the monochromatization principle has never been thoroughly investigated to the point of actual implementation in a collider for experimental validation. Impediments to implementing the various proposals included: insufficient studies and uncertainty of performance; the cost and effort of hardware implementation; shifting perceptions of the physics interest, which resulted in significant performance improvements at the operating colliders without the introduction of monochromatization.

Since the first proposal of monochromatization for the FCC-ee in 2016~\cite{ValdiviaGarcia:2016rdg}, recent studies have revisited and extended the concept of monochromatization in the context of circular $e^+ e^-$ colliders at high energies and in crossing-angle collision configuration~\cite{ValdiviaGarcia:2016rdg,ValdiviaGarcia:2016fae,ValdiviaGarcia:2017xiv,Bogomyagkov:2017tpk,Shatilov:2018drj,Zimmermann:2017tjv,ValdiviaGarcia:2018cpm,ValdiviaGarcia:2019ezi,telnov2020monochromatizationeecolliderslarge,Garcia:2019nci,ValdiviaGarcia:2022nks,Faus-Golfe:2021udx}. Under these conditions, two new issues need to be addressed.
\begin{figure}[h!]
\centering
\includegraphics[width=0.5\textwidth]{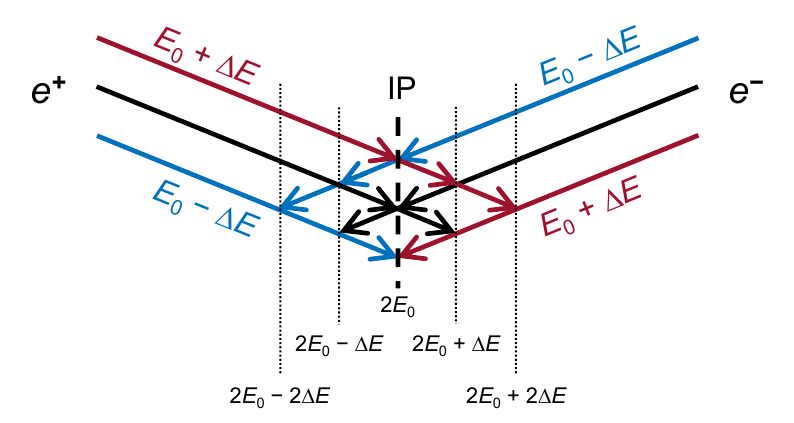}
\caption{Crossing-angle collision scheme with monochromatization.}\label{fig2}
\end{figure}

The first involves extending the previous monochromatization concept, originally proposed for head-on collisions, to configuration with a crossing angle. Although the nonzero collision angle ($\theta_{c}$) decreases monochromatization efficiency due to the correlation between the collision energy and longitudinal position, as illustrated in Fig.~\ref{fig2}, it remains feasible since particles with energy deviations still encounter a higher density of particles with opposite energy deviations. With nonzero $\theta_{c}$ in the horizontal plane, the $\mathcal{L}$ and beam-beam tune shift ($\xi_{x,y}$) are supposed to be corrected by replacing the $\sigma_{x}^{\ast}$ with the effective beam size $\sigma_{x}^{\ast} \sqrt{1+\varphi^2}$~\cite{sen2022luminositybeambeamtuneshifts}. The Piwinski angle ($\varphi$) is given by: $\varphi = (\sigma_{z}/\sigma_{x}^{\ast})\tan (\theta_{c}/2)$ where $\sigma_{z}$ is the bunch length~\cite{Raimondi:2007vi}. The $\mathcal{L}$ still follows the relation described in Eq.~\ref{eq2} with the extended $\lambda$ defined as:
\begin{equation}
    \lambda = \sqrt{1 + \sigma_{\delta}^2 \left( \frac{D_x^{\ast 2}}{\varepsilon_{x} \beta_{x}^{\ast} (1+\varphi^2)} + \frac{D_y^{\ast 2}}{\varepsilon_{y} \beta_{y}^{\ast}} \right) }\:.
\label{eq3}
\end{equation}
The reduced $\sigma_{W}$ is defined as:
\begin{equation}
    \sigma_{W} = \sqrt{2} E_0 \sqrt{\left(\frac{\sigma_{\delta}\cos(\theta_{c}/2)}{\lambda}\right)^2 + \left(\sigma_{x^\prime}^{\ast}\sin(\theta_{c}/2)\right)^2}
\label{eq4}
\end{equation}
with $\sigma_{x^{\prime}}^{\ast} = \sqrt{\varepsilon_{x}/\beta_{x}^{\ast}}$ the beam angular spread at the IP in the horizontal plane~\cite{Bogomyagkov:2017tpk}. The luminosity loss still follows the previously established relationship described in Eq.~\ref{eq2}.

The second issue arises from the high energies. In existing low-energy circular $e^+ e^-$ colliders, $\sigma_{\delta}$ of the beams is mainly due to SR emitted when a charged ultra-relativistic particle passes through the accelerator bending magnets (rings). In collider physics, the SR emitted due to the electromagnetic field of the opposing beam during collision, is known as beamstrahlung (BS). This BS effect is negligible in the current low-energy circular $e^+ e^-$ colliders. On the contrary, in future high-energy circular $e^+ e^-$ colliders, it will become of significant magnitude for the first time, contributing to an increase of $\sigma_{\delta}$ and impacting $\varepsilon_{x,y}$ and $\sigma_{z}$. Introducing the usual equilibrium parameters, determined by the arc SR alone, with the subindex “SR,” and the total values in collision, including the effect of BS, with the subscript “tot,” and with the hypothesis that the strength of SR characterized by $\Upsilon = \hbar \omega_{c}/E_{e} \ll 1$ ($\omega_{c}$ the critical energy and $E_{e}$ the electron energy before radiation) and $\sigma_{x}^{\ast} \gg \sigma_{y}^{\ast}$ (vertical flat beams), the $\sigma_{\delta,\text{tot}}$ and the $\varepsilon_{x,y,\text{tot}}$ with the BS impact are given by the following coupled equations~\cite{Garcia:2019nci,ValdiviaGarcia:2022nks}:
\begin{equation}
    \sigma_{\delta,\text{tot}}^2 = \sigma_{\delta,\text{SR}}^2 + \frac{V}{\sigma_{\delta,\text{tot}}^2 \sigma_{x,\text{tot}}^{\ast 3}}\:,
\label{eq5}
\end{equation}
\begin{equation}
    \varepsilon_{x,\text{tot}} = \varepsilon_{x,\text{SR}} + \frac{2V \mathcal{H}_{x}^{\ast}}{\sigma_{\delta,\text{tot}}^2 \sigma_{x,\text{tot}}^{\ast 3}}\:,
\label{eq6}
\end{equation}
\begin{equation}
    \varepsilon_{y,\text{tot}} = \varepsilon_{y,\text{SR}} + \frac{2V \mathcal{H}_{y}^{\ast}}{\sigma_{\delta,\text{tot}}^2 \sigma_{x,\text{tot}}^{\ast 3}}
\label{eq7}
\end{equation}
with $\sigma_{\delta,\text{SR}}$ and $\varepsilon_{x,y,\text{SR}}$ the SR contributions to the $\sigma_{\delta}$ and $\varepsilon_{x,y}$, respectively, $\sigma_{x,\text{tot}}^{\ast} = \sqrt{\varepsilon_{x,\text{tot}}\beta_{x}^{\ast}+D_{x}^{\ast 2} \sigma_{\delta,\text{tot}}^2}$ the total horizontal beam size and $\mathcal{H}_{x,y}^{\ast}$ the IP dispersion invariant given by:
\begin{equation}
    \mathcal{H}_{x,y}^{\ast} = \frac{\left( \beta_{x,y}^{\ast} D_{x,y}^{\prime \ast} + \alpha_{x,y}^{\ast} D_{x,y}^{\ast} \right)^2 + \left( D_{x,y}^{\ast} \right)^2}{\beta_{x,y}^{\ast}}\:.
\label{eq8}
\end{equation}
For head-on collisions without $\theta_{c}$, the $V$ factor is approximated as:
\begin{equation}
    V \approx \frac{55 \pi^2}{3 \sqrt{3}} \left(\sqrt{\frac{2}{\pi}}\right)^3 \frac{n_{\text{IP}} \tau_{E,\text{SR}} Q_{s}^2 r_{e}^5 \gamma^2 N_{b}^3}{T_{\text{rev}} \alpha_{\text{C}}^2 C^2 \alpha} \times 0.7183\:,
\label{eq9}
\end{equation}
where $n_{\text{IP}}$ denotes the number of IPs, $\tau_{E,\text{SR}}$ the longitudinal damping time, $Q_{s}$ the synchrotron tune, $r_{e}$ the classical electron radius, $\gamma$ the relativistic Lorentz factor, $N_{b}$ the bunch population, $T_{\text{rev}}$ the revolution time, $\alpha_{\text{C}}$ the momentum compaction factor, $C$ the circumference and $\alpha$ the fine structure constant. In crossing-angle collisions, $\varphi$ needs to be introduced and we can obtain an approximation of the $V$ factor as:
\begin{equation}
    V \approx \frac{55 \pi^2}{3 \sqrt{3}} \left(\sqrt{\frac{2}{\pi}}\right)^3 \frac{n_{\text{IP}} \tau_{E,\text{SR}} Q_{s}^2 r_{e}^5 \gamma^2 N_{b}^3}{T_{\text{rev}} \alpha_{\text{C}}^2 C^2 \alpha} \times \frac{0.77562}{\varphi}\:.
\label{eq10}
\end{equation}
Notice that $\varphi$ in this $V$ factor should be calculated using the $\sigma_{x,\text{tot}}^{\ast}$ and the total bunch length ($\sigma_{z,\text{tot}}$). In this context, the $\sigma_{z,\text{tot}}$ is always determined by:
\begin{equation}
    \sigma_{z,\text{tot}} = \frac{\alpha_{\text{C}}C}{2 \pi Q_{s}} \sigma_{\delta,\text{tot}}\:.
\label{eq11}
\end{equation}
As the $\sigma_{\delta,\text{tot}}$ and $\varepsilon_{x,y,\text{tot}}$ are coupled in a self-consistency condition, Eq.~\ref{eq5}, Eq.~\ref{eq6} and Eq.~\ref{eq7} need to be solved together for the $\sigma_{\delta,\text{tot}}$ and $\varepsilon_{x,y,\text{tot}}$.

Considering the impact of BS, comprehensive self-consistent parametric studies of FCC-ee with monochromatized head-on IP beams at \textit{s}-channel Higgs production energy ($\sim$\SI{125}{\giga\electronvolt}) have been performed to identify the best scenario with $D_{x}^{\ast} \neq 0$~\cite{ValdiviaGarcia:2016rdg,ValdiviaGarcia:2017xiv,Zimmermann:2017tjv,ValdiviaGarcia:2022nks,Faus-Golfe:2021udx}. As previously mentioned, at this high energy the natural $\sigma_{W}$ due to SR is $\sim$\SI{50}{\mega\electronvolt}. In standard FCC-ee running conditions without $D_{x}^{\ast}$, the $\sigma_{W}$ blows up to $\sim$\SI{70}{\mega\electronvolt} due to BS from Eq.~\ref{eq5}, while $\varepsilon_{x,\text{tot}} \approx \varepsilon_{x,\text{SR}}$ with zero $\mathcal{H}_{x}^{\ast}$ ($D_{x}^{\ast} = 0,\,D_{x}^{\prime \ast} = 0,\,\alpha_{x}^{\ast} = 0$) in Eq.~\ref{eq6}. The $\sigma_{W}$ can be reduced, by up to one order of magnitude ($\lambda \approx 10$), by means of monochromatization~\cite{Renieri:1975wt,Bassetti:1974if,Protopopov:1979tn,Avdienko:1983mee,Wille:1984fb,Jowett:1985zx,Bassetti:1987zg,Jowett:806265,DUBROVIN:1991,Veshcherevich:1992by,Alexahin:1990zps,Zholents:1992ua,Faus-Golfe:1996uai,Schulte:1997nga,Zholents:1988bu}. If we suppose a nonzero $D_{x}^{\ast}$ of opposite sign for the two colliding beams, this nonzero $D_{x}^{\ast}$ leads to a significant $\varepsilon_{x}$ blow-up due to BS from Eq.~\ref{eq6} with nonzero $\mathcal{H}_{x}^{\ast} = D_{x}^{\ast 2}/\beta_{x}^{\ast}$ ($D_{x}^{\prime \ast} = 0,\,\alpha_{x}^{\ast} = 0$). At the same time, thanks to the much-increased $\sigma_{x}^{\ast}$, the $\sigma_{\delta}$ and $\sigma_{z}$ return to their natural values without collision, from Eq.~\ref{eq5}. This effect alone already decreases the collision energy spread. An additional reduction is caused by colliding particles with exactly opposite energy deviations, thanks to the opposite sign of $D_{x}^{\ast}$. Notice that simulations in Refs.~\cite{ValdiviaGarcia:2016rdg,ValdiviaGarcia:2017xiv,Zimmermann:2017tjv,ValdiviaGarcia:2022nks,Faus-Golfe:2021udx} supposed ideal Gaussian distribution of particles at the IP, but not the associated optics to generate them.

Another factor that could have an impact on this collision scheme is the hourglass effect~\cite{Bogomyagkov:2017tpk,Shatilov:2018drj}. This effect reduces the $\mathcal{L}$ by a reduction factor ($R_{\text{hg}}$) that is estimated at 0.3 based on the self-consistent parameters for monochromatization at the FCC-ee~\cite{ValdiviaGarcia:2022nks,Faus-Golfe:2021udx}. This factor can potentially be further optimized by shaping the $\beta_{y}^{\ast}$. For instance, setting $\beta_{y}^{\ast}$ to \SI{4}{\milli\meter} could increase $R_{\text{hg}}$ to $\sim$0.8. In this paper, this effect is not addressed, but it remains feasible for optimization if necessary.

The purpose of this paper is to report on the developments of realistic IR optics designs for monochromatization at the FCC-ee, improving upon the previous self-consistent parametric studies~\cite{ValdiviaGarcia:2022nks,Faus-Golfe:2021udx}. In the following sections, we provide a detailed description of the optimized IR optics design for different possible monochromatization scenarios, along with their performance evaluations after implementation into the FCC-ee global lattice, including realistic beam dynamic simulations and estimates of the best physics working points (in terms of $\sigma_{W}$ and $\mathcal{L}_{\text{int}}$) aiming to measure the electron Yukawa coupling. Considering that monochromatization has never been tested experimentally before, all the proposed new IR optics have been designed to be compatible with a standard operation mode, featuring no dispersion at the IP, to mitigate potential risks.

\section{The baseline FCC-ee GHC lattice and new ideas for implementing the monochromatization operation mode}
\label{sec2}
The layout of the FCC-ee baseline consists of two horizontally separated rings for $e^+$ and $e^-$ and follows a low-risk tunnel scenario, with a circumference of approximately \SI{91}{\kilo\meter} and eight straight sections. The baseline FCC-ee standard lattice design is the so-called “Global Hybrid Correction” (GHC) optics~\cite{Oide:2016mkm,Keintzel:2023myn,vanRiesen-Haupt:2024ore}, designed to operate at four different beam energies of 45.6, 80, 120, and \SI{182.5}{\giga\electronvolt}, allowing physics precision experiments at the Z-pole (Z mode), the W-pair-threshold (WW mode), the ZH-maximum (ZH mode) and above the top-pair-threshold ($\mathrm{t\bar{t}}$ mode), respectively. It provides four experimental IRs where the $e^+$ and $e^-$ beams are brought to collision from the inside outwards with a $\theta_{c}$ of \SI{30}{\milli\radian} in the horizontal plane and a vertical crab-waist scheme. This scheme consists of a local chromaticity correction (LCC) system implemented only in the vertical plane, resulting in nonzero horizontal dispersion ($D_{x}$) created by weak horizontal dipole magnets, disposed in an asymmetric configuration at the two sides of each IP to avoid SR backgrounds in the IP region, where the vertical chromaticity correction sextupole pairs are located.

Given the presence of horizontal bending magnets in the LCC system, the most natural way to implement monochromatization in this FCC-ee lattice type is reconfiguring these LCC horizontal dipoles to generate a nonzero $D_{x}^{\ast}$, while maintaining the same $\theta_{c}$. Indeed, from Eq.~\ref{eq5}, a wide $\sigma_{x}^{\ast}$ helps to mitigate the impact of BS on $\sigma_{\delta}$, while preserving a small $\sigma_{y}^{\ast}$ is crucial for attaining a high $\mathcal{L}$. Taking into account the reference parameter table~\cite{Keintzel:2023myn} for the FCC-ee GHC lattices with horizontal betatron sizes ($\sigma_{x\beta}^{\ast} = \sqrt{\varepsilon_{x} \beta_{x}^{\ast}}$) at the IP of the order of \si{\milli\meter} and a $\sigma_{\delta,\text{SR}}$ of $\sim$0.05\% at \textit{s}-channel Higgs production energy ($\sim$\SI{125}{\giga\electronvolt}), a $D_{x}^{\ast}$ of around \SI{10}{\centi\meter} is required to achieve a $\lambda$ of $\sim$\SIrange{5}{8}{} from Eq.~\ref{eq3}.

Because $\sigma_{y\beta}^{\ast}$ ($\sim$\si{\nano\meter}) $\ll$ $\sigma_{x\beta}^{\ast}$ ($\sim$\si{\micro\meter}) for getting high luminosities, about 100 times smaller $D_{y}^{\ast}$ ($\sim$\si{\milli\meter}) is needed to get a similar $\lambda$. Thus, a nonzero $D_{y}^{\ast}$ of this magnitude could be generated by simply using skew quadrupole correctors around the IP. These quadrupoles could be located close to the sextupole pairs within the LCC system.

The monochromatization IR optics designs presented in the following sections are based on the Version 2022 of FCC-ee GHC optics (FCC-ee GHC V22)~\cite{fcc_lattice_v22}. There are two types of FCC-ee GHC V22 lattices specifically designed for the Z mode and the $\mathrm{t\bar{t}}$ mode. Apart of the $E_0$ and the beam current ($I$), from the optics point of view these lattices differs on the $\beta_{x,y}^{\ast}$ and the arc cells type: Long 90/90 for the Z mode and 90/90 for the $\mathrm{t\bar{t}}$ mode. For each of these two baseline lattice types, labeled as “FCC-ee GHC V22 Z” and “FCC-ee GHC V22 $\mathrm{t\bar{t}}$,” respectively, we implemented three monochromatization scenarios with $D_{x}^{\ast}$ only, $D_{y}^{\ast}$ only, and combined $D_{x}^{\ast}$ and $D_{y}^{\ast}$. Furthermore, we matched a standard optics configuration ($D_{x}^{\ast} = 0$) with an orbit path similar to that of the monochromatization orbit in the case where we introduced $D_{x}^{\ast} \neq 0$.

\section{FCC-ee GHC monochromatization IR optics design with horizontal IP dispersion}
\label{sec3}

\subsection{Monochromatization IR optics design}
\label{subsec3.1}
As mentioned in the previous section, the generation of the $D_{x}^{\ast}$ could be achieved by re-parameterizing the angle of the LCC horizontal dipoles at each side of the IP, without changes in their position or their length. The monochromatization IR optics design based on each type of baseline standard optics (“FCC-ee GHC V22 Z” and “FCC-ee GHC V22 $\mathrm{t\bar{t}}$”) was conducted in steps using the MAD-X program~\cite{madx}, with the target IP parameters taken from the monochromatization self-consistent parameter optimization: $\beta_{x,y}^{\ast} = 90,\SI{1}{\milli\meter}$ with $D_{x}^{\ast} = \SI{0.105}{\meter}$~\cite{ValdiviaGarcia:2022nks,Faus-Golfe:2021udx}.

\begin{figure}[h!]
\centering
\includegraphics[width=0.5\textwidth]{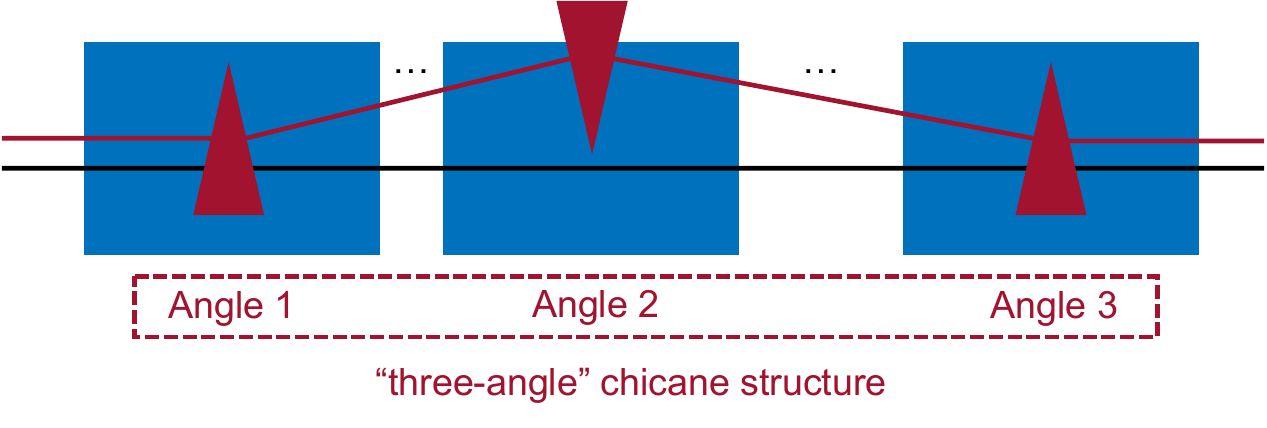}
\caption{Layout of the “three-angle” chicane structure. The LCC horizontal dipoles are depicted in blue, while the angles of the chicane structure are in red. The baseline standard orbit is shown in black, and the monochromatization orbit in red.}\label{fig3}
\end{figure}

First, the dispersion in the LCC region was deliberately “mismatched” to generate the required $D_{x}^{\ast}$. We modified the angle of the last three upstream and downstream horizontal dipoles of the LCC region to match the dispersion back to zero outside of the LCC region, following a chicane structure with three angles, as schematically depicted in Fig.~\ref{fig3}, which is so-called the “three-angle” chicane structure. In this configuration, where $L_{12}$ is the distance from Angle 1 to Angle 2 and $L_{23}$ is the distance from Angle 2 to Angle 3, the ratio of the values of Angle 1 and Angle 3 exhibits an inverse proportionality with respect to the ratio of $L_{12}$ and $L_{23}$: $\theta_1 = (L_{23}/L_{12})\theta_3$. The value of Angle 2 is opposite in sign and equal in magnitude to the sum of $\theta_1$ and $\theta_3$: $\theta_2=-(\theta_1+\theta_3)$. The orbit of the monochromatization IR with this close orbit chicane structure differs from the standard one only at the locations where these chicanes are implemented. As a result, the positions and angles at the entrance and exit of the monochromatization IR orbits remain identical to those of the standard orbit. To enhance the flexibility and optimize the matching efficiency in MAD-X, all the LCC horizontal dipoles were split into three sets, with additional thin quadrupoles inserted between these split sets~\cite{Jiang:2022oyt,Zhang:2023lae,Zhang:2023cepc}. Subsequently, the $\beta_{x,y}^{\ast}$ were matched to meet the target values, while ensuring that the beam parameters at the entrance and exit of the IR were consistent with the standard IR optics. Additionally, the phase advances between the IP and the LCC sextupole pairs were matched to those in the standard IR optics to facilitate further local chromaticity corrections.

\begin{figure}[h!]
\centering
\includegraphics[width=0.5\textwidth]{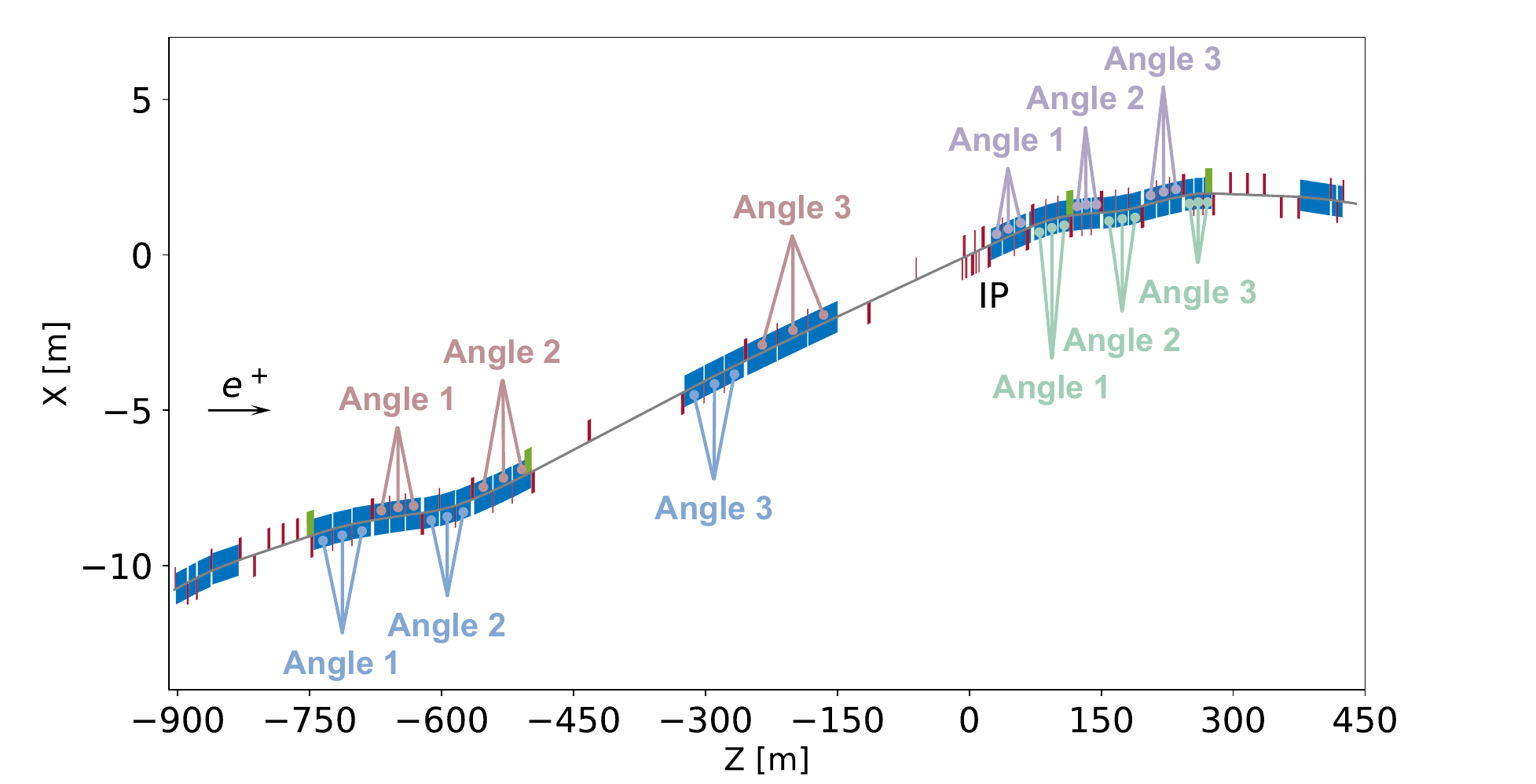}
\caption{Optimized monochromatization orbit with four sets of smooth close orbit “three-angle” chicane structure indicated by different colors. The beam direction is from left to right and the point $\text{X}=0$, $\text{Z}=0$ marks the location of IP. Dipoles, quadrupoles, and sextupoles are shown in blue, red, and green, respectively, while focusing and defocusing elements are positioned above and below the orbit.}\label{fig4}
\end{figure}

However, we observed that the angles of added chicane structures were significantly larger than those of the original LCC horizontal dipoles. This resulted in a high level of SR and a blow-up of $\varepsilon_{x}$~\cite{Zhang:2023cepc}. To mitigate this blow-up, we implemented two longer “three-angle” chicane structures to both the upstream and downstream IR, resulting in a total of four chicane structures being added to the twelve LCC horizontal dipoles within the IR. Each chicane angle was equally distributed among the three split dipole sets. After performing step-by-step matching and optimization in MAD-X, we ultimately integrated four sets of smooth close orbit “three-angle” chicane structures into the LCC system~\cite{Zhang:2024cepceu,Zhang:2024uxh,Fausgolfe:2024ipac,Zhang:2024phd}. Fig.~\ref{fig4} shows the optimized monochromatization orbit based on “FCC-ee GHC V22 Z” baseline standard optics.

The resulting monochromatization IR lattice and optics with a $D_{x}^{\ast}$ of \SI{0.105}{\meter} based on “FCC-ee GHC V22 Z” optics are presented in the top panel of Fig.~\ref{fig5}. In addition to previously mentioned constraints on the beam parameters, the four peak values of the $\beta_{y}$ at the sextupole pair locations were matched to be the same, while $D_{x}$ at the positions of the inner sextupole pairs was kept nonzero to reduce the required sextupole strengths. Furthermore, the dispersion-free regions before the arcs are preserved, maintaining consistency with the original standard optics configuration.

\subsection{Orbit compatibility with standard operation mode}
\label{subsec3.2}
\begin{figure}[h!]
\centering
  \begin{subfigure}[b]{0.5\textwidth}
     \includegraphics[width=\textwidth]{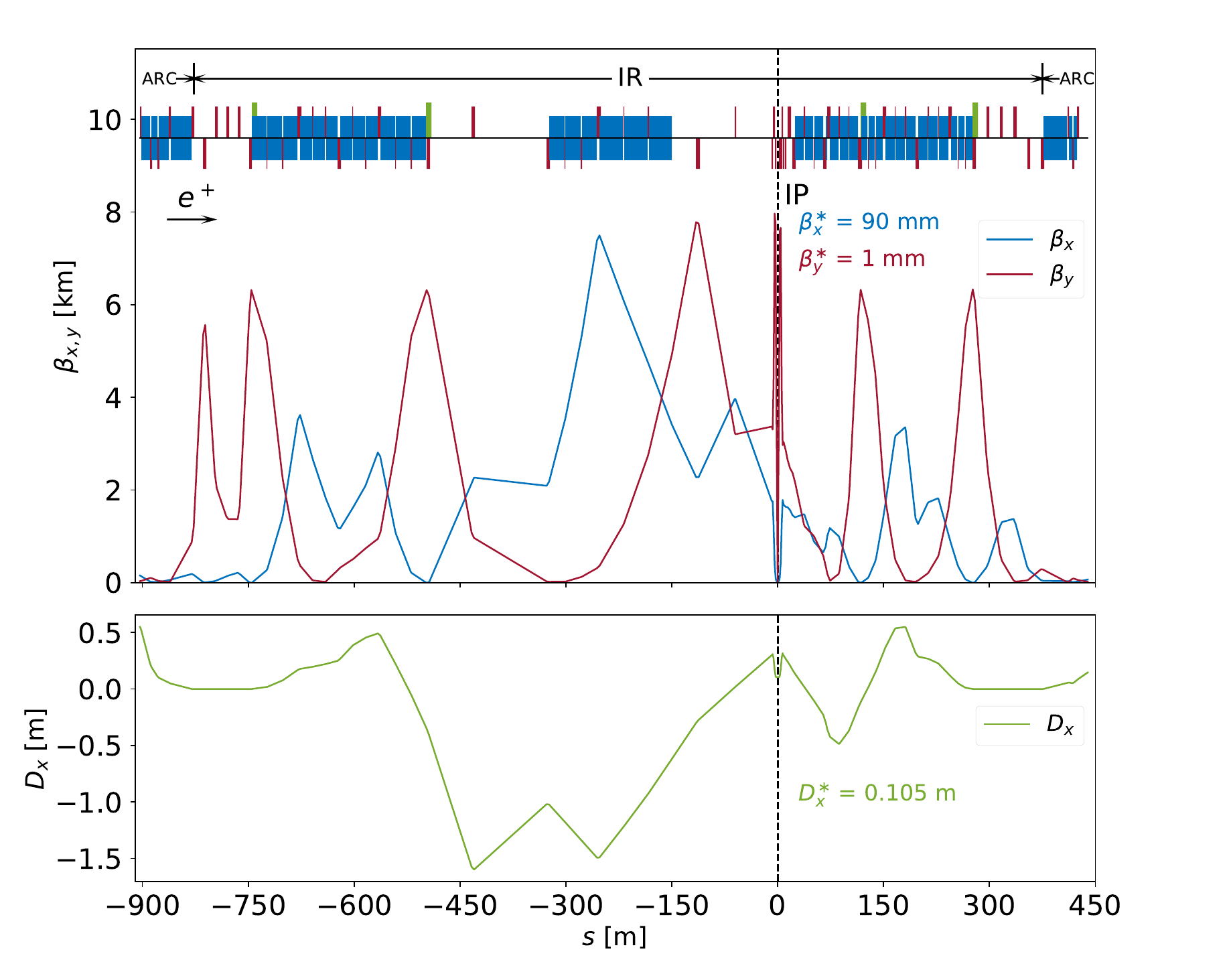}
  \end{subfigure}
  \begin{subfigure}[b]{0.5\textwidth}
     \includegraphics[width=\textwidth]{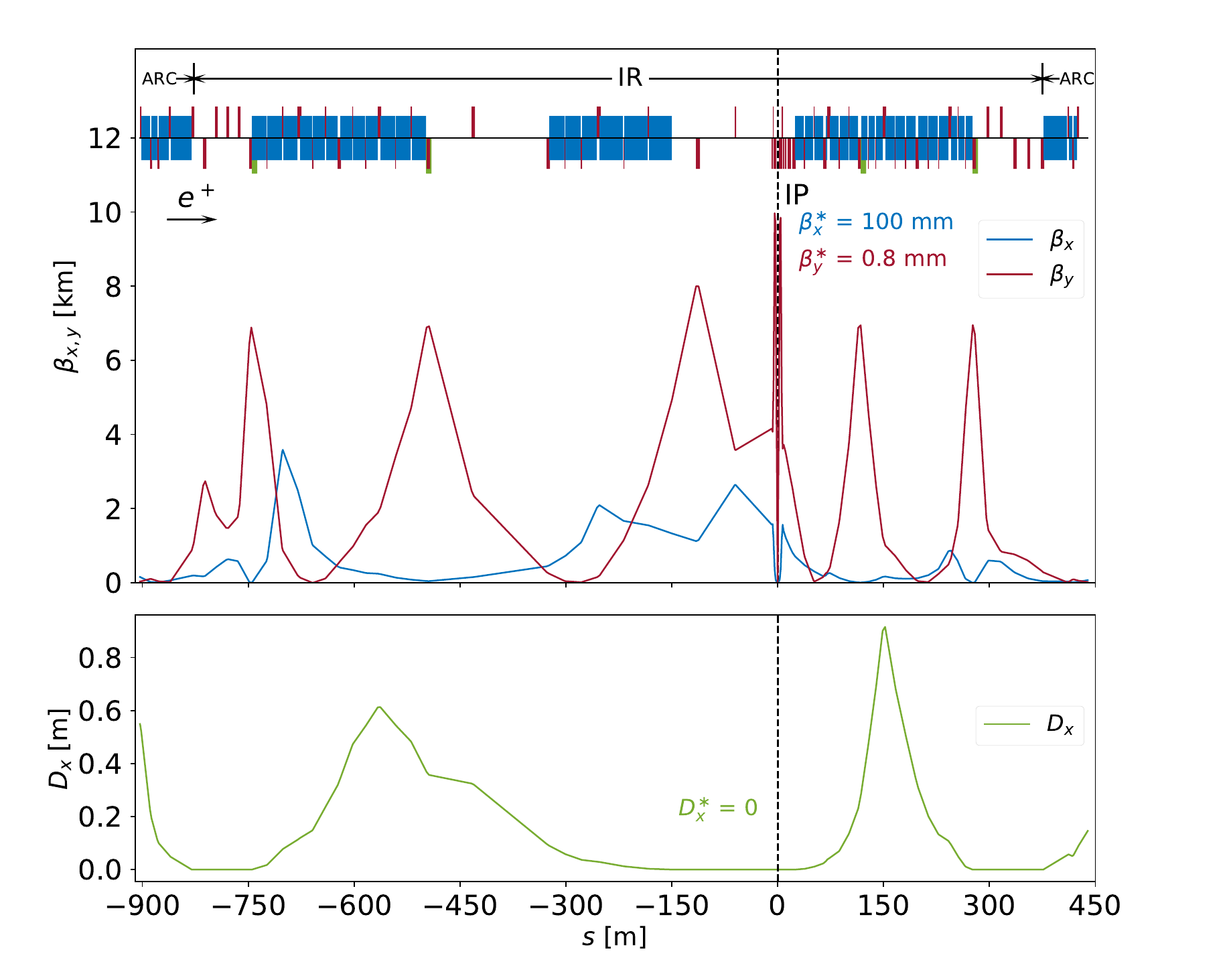}
  \end{subfigure}
\caption{Lattices and optics of the monochromatization IR with $D_{x}^{\ast}$ of \SI{0.105}{\meter} (top) and the corresponding standard mode IR with orbit compatibility (bottom) based on the “FCC-ee GHC V22 Z” optics calculated using MAD-X. The beam direction is from left to right; the dashed line at $s=0$ marks the IP. In the lattice, dipoles, quadrupoles, and sextupoles are shown in blue, red, and green, respectively, while focusing and defocusing elements are positioned above and below the orbit. In the optics, $\beta_{x,y}$ are shown in blue and red, respectively, while $D_{x}$ is shown in green.}\label{fig5}
\end{figure}

\begin{figure}[h!]
\centering
\includegraphics[width=0.5\textwidth]{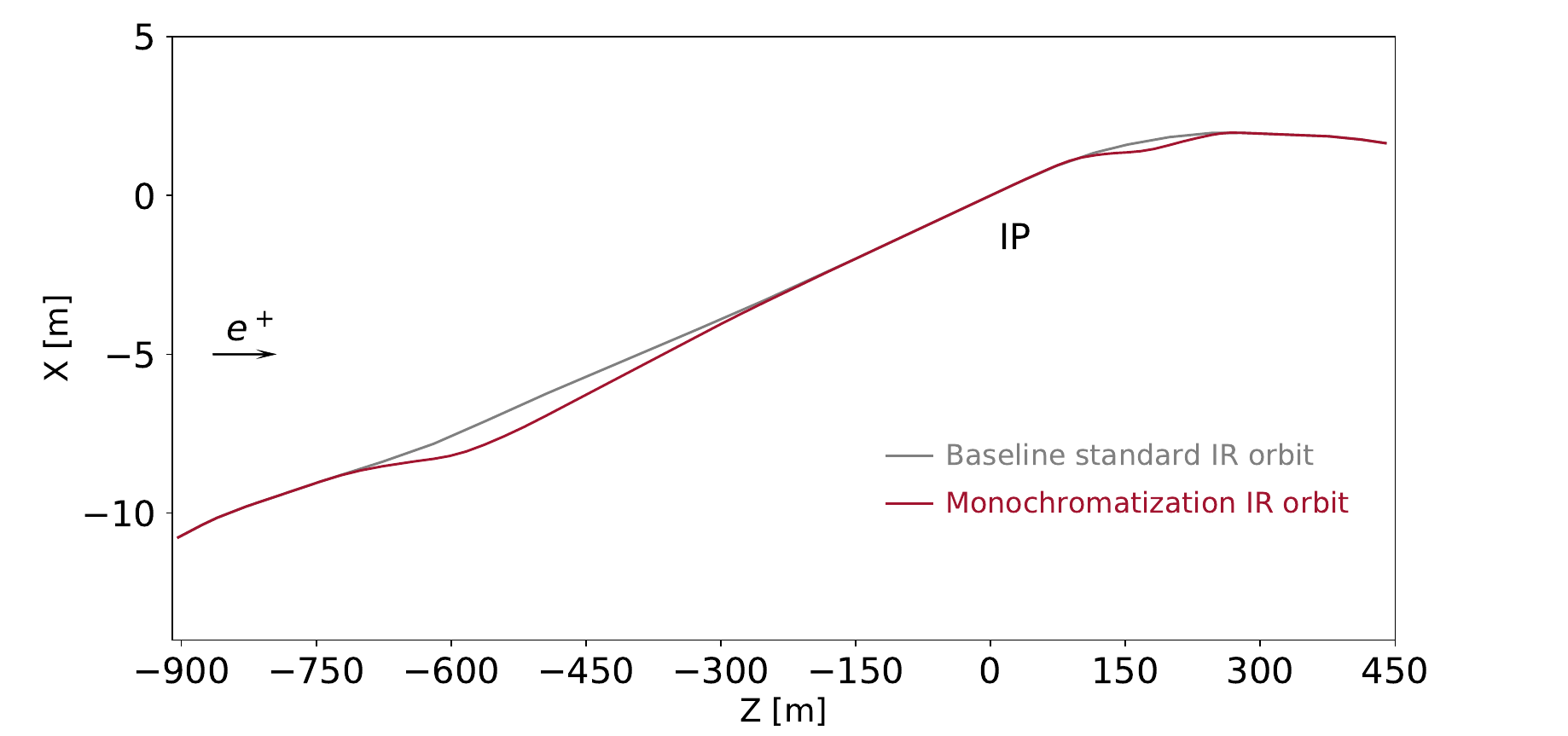}
\caption{Comparison of the “FCC-ee GHC V22 Z” IR orbit (grey) and its monochromatization IR orbit (red) with four sets of smooth close orbit “three-angle” chicane structure. The beam direction is from left to right; the point $\text{X}=0$, $\text{Z}=0$ marks the IP location.}\label{fig6}
\end{figure}

The comparison between the “FCC-ee GHC V22 Z” IR orbit and its corresponding monochromatization IR orbit is shown in Fig.~\ref{fig6}. As the orbit of the monochromatization IR is slightly different from the original IR orbit, it becomes necessary to rematch the standard optics without IP dispersion using the monochromatization orbit (incorporating chicane structures) to enable further comparison and demonstrate orbit compatibility with the standard operation mode. By fixing the angles of all horizontal dipoles within the IR, thus preserving the monochromatization orbit, we re-matched the beam parameters at the IP to align with those of the original optics. The result of this process is presented in the bottom panel of Fig.~\ref{fig5}.

\begin{figure}[h!]
\centering
  \begin{subfigure}[b]{0.5\textwidth}
     \includegraphics[width=\textwidth]{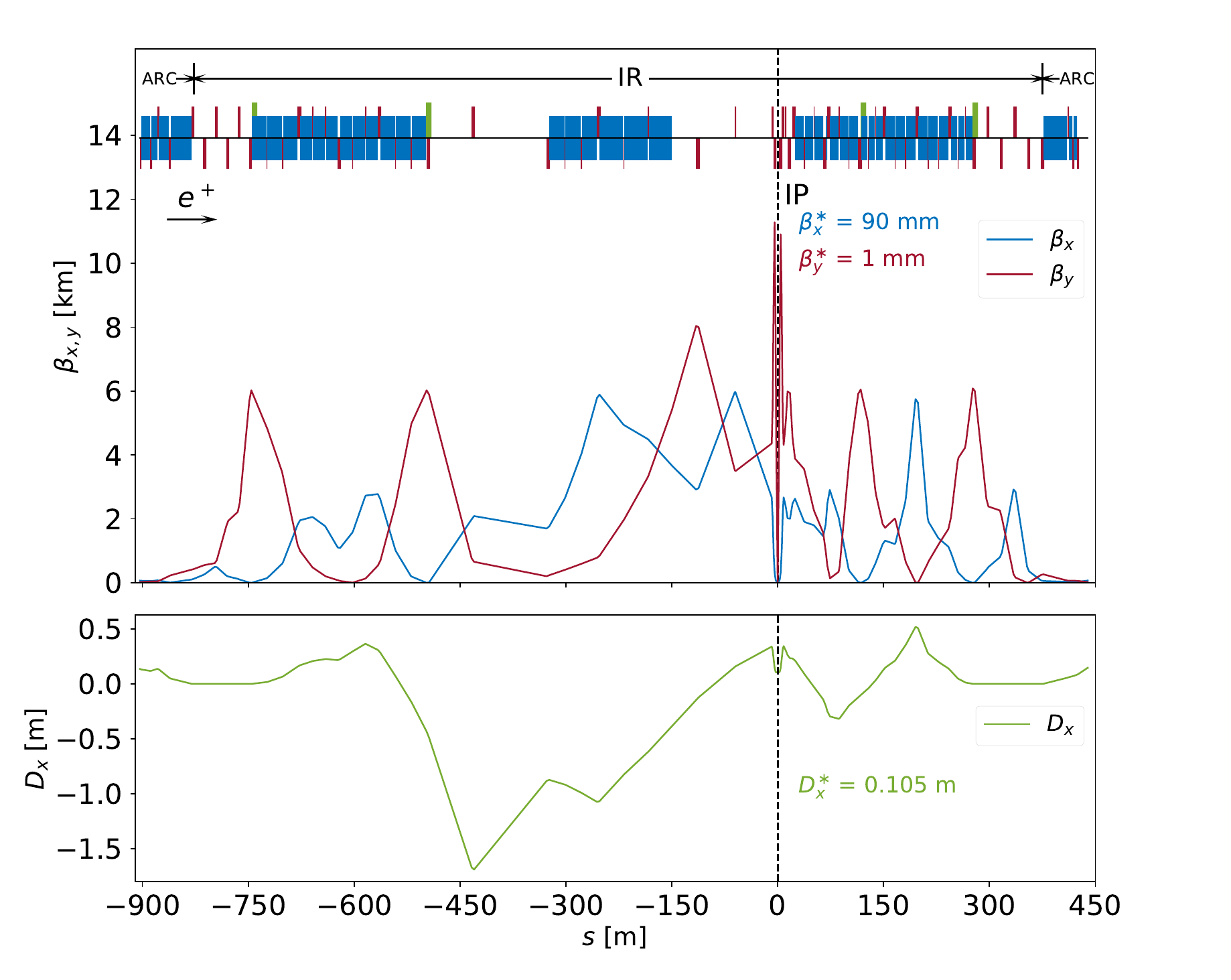}
  \end{subfigure}
  \begin{subfigure}[b]{0.5\textwidth}
     \includegraphics[width=\textwidth]{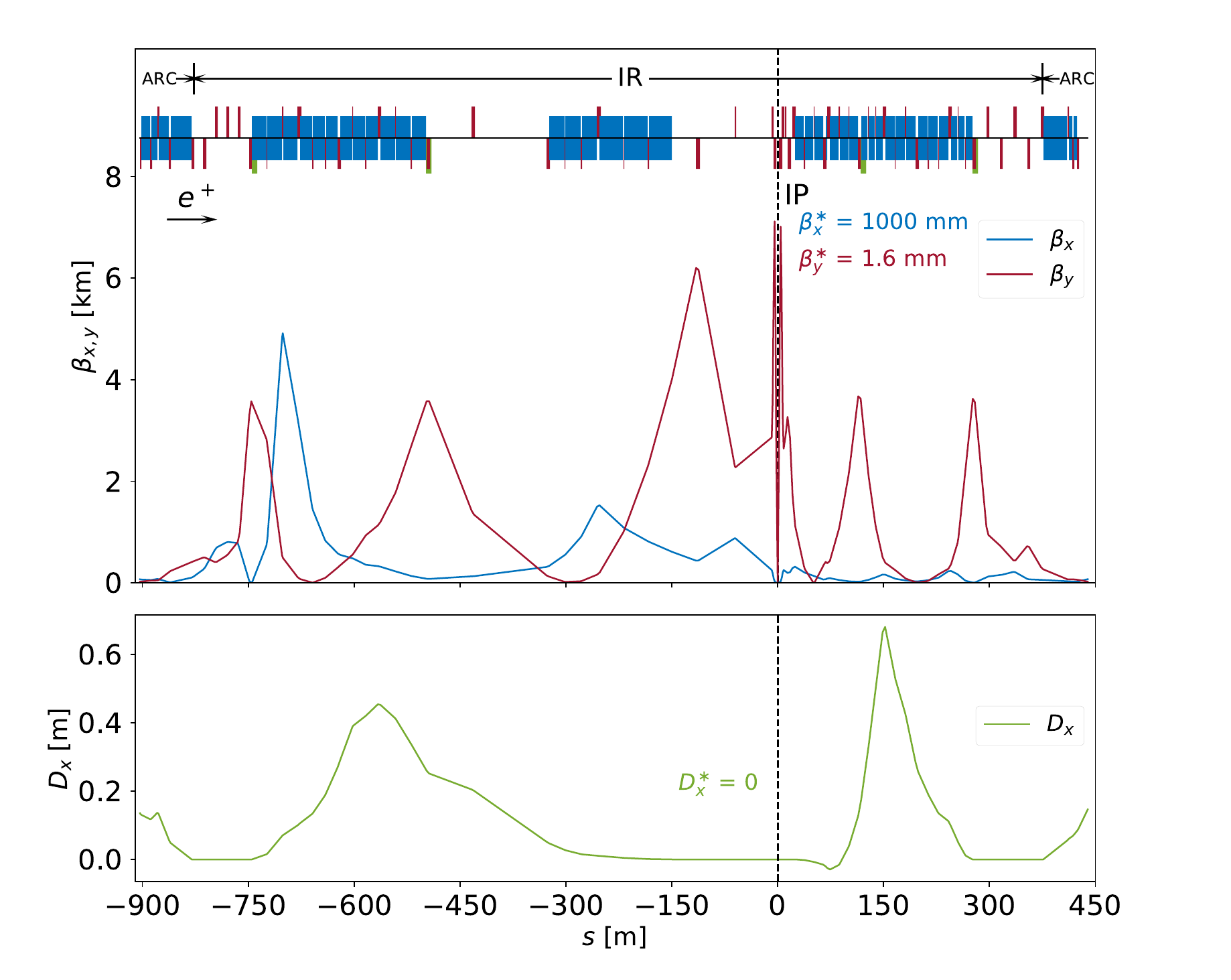}
  \end{subfigure}
\caption{Lattices and optics of the monochromatization IR with $D_{x}^{\ast}$ of \SI{0.105}{\meter} (top) and the corresponding standard mode IR with orbit compatibility (bottom) based on the “FCC-ee GHC V22 $\mathrm{t\bar{t}}$” optics calculated using MAD-X. The beam direction is from left to right; the dashed line at $s=0$ marks the IP. In the lattice, dipoles, quadrupoles, and sextupoles are shown in blue, red, and green, respectively, while focusing and defocusing elements are positioned above and below the orbit. In the optics, $\beta_{x,y}$ are shown in blue and red, respectively, while $D_{x}$ is shown in green.}\label{fig7}
\end{figure}

A similar approach was also applied to the “FCC-ee GHC V22 $\mathrm{t\bar{t}}$” optics~\cite{Zhang:2024cepceu,Zhang:2024uxh,Fausgolfe:2024ipac,Zhang:2024phd}. Fig.~\ref{fig7} shows the monochromatization IR optics with a $D_{x}^{\ast}$ of \SI{0.105}{\meter} at the top, and the corresponding standard mode IR optics with orbit compatibility at the bottom.

\subsection{Global optical performance calculation}
\label{subsec3.3}
With the same beam parameters and orbit at the entrance and exit of the monochromatization IR as the standard one, we implemented the new IR into the global lattice as a replacement for the original standard IR and performed the following steps before the global optical performance calculation~\cite{Zhang:2023cepc,Zhang:2024cepceu,Fausgolfe:2024ipac,Zhang:2024phd}. First, to control the chromaticity generated within the IR, the FCC-ee LCC system consists of non-interleaved sextupole pairs integrated for the vertical plane on both sides of the IP. The outer sextupole pairs, known as crab sextupoles (CS), are also used for crab-waist transformation~\cite{Zobov:2016sxm}. After matching the vertical chromaticity from the IP to CS pairs to zero by varying the strength of the inner chromatic sextupoles $k_{2,\text{chrom}}$, the strength of the CS $k_{2,\text{CS}}$ follows from~\cite{Oide:2016mkm}:
\begin{equation}
    k_{2,\text{CS}} = k_{2,\text{chrom}} - \frac{F}{\theta_{c}\beta_{y}^{\ast}\beta_{y}}\sqrt{\frac{\beta_{x}^{\ast}}{\beta_{x}}}
\label{eq12}
\end{equation}
with $\beta_{x,y}$ the betatron function at the locations of CS pairs. The crab factor $F$ is determined from beam-beam studies, with values of 97\% for the Z mode and 87\% for the WW mode. We temporarily set it at 90\% for our monochromatization optics. Secondly, for the global chromaticity correction, the strength of all focusing and defocusing sextupoles was multiplied by two respective coefficients, respectively. By varying these two coefficients, the global horizontal and vertical chromaticities were matched to a value of 5, as positive global chromaticity improves beam stability. Next, by adjusting the strengths of the quadrupoles in the RF insertions, we corrected both the horizontal and vertical tune to be consistent with those of the standard optics, while maintaining the beam parameters in the IRs. Finally, we switched on the RF cavities to compensate for the energy loss due to SR. The value of the energy difference, divided by the reference momentum times the velocity of light: $t^{\prime} = \Delta E/pc$, was matched to zero by varying the phase of the RF cavities. The voltage of the RF cavity ($V_{\text{RF}}$) affects the $Q_{s}$ according to the relation:
\begin{equation}
    Q_{s}^2 = \frac{\alpha_{\text{C}}h}{2\pi E_0} \sqrt{e^2 V_{\text{RF}}^2-U_0^2}
\label{eq13}
\end{equation}
with $h$ the harmonic number, $e$ the elementary charge and $U_0$ the energy loss per turn. This, in turn, influences the $\sigma_{z}$ as described in Eq.~\ref{eq11}. Therefore, an appropriate $V_{\text{RF}}$ was selected to achieve a $\sigma_{z}$ of $\sim$\SI{4}{\milli\meter}, which falls between the $\sigma_{z}$ of the Z mode and WW mode.

After applying all corrections and SR-RF strategy compensation, the values of $U_0$, $\alpha_{\text{C}}$, $Q_{s}$, $\tau_{E,\text{SR}}$, $\sigma_{\delta}$, $\varepsilon_{x}$, and $\sigma_{z}$ were calculated using MAD-X, considering only the SR. The $\varepsilon_{y}$ was calculated with the assumed FCC-ee coupling ratio (full current) of motion between the horizontal and vertical planes: $\kappa = \varepsilon_{y}/\varepsilon_{x} = 0.2\%$, which can be matched by installing wigglers in the arcs. The $I$ was maximized based on the maximum allowable total synchrotron radiation power loss ($P_{\text{SR}}$) of \SI{50}{\mega\watt}, following the relation: $P_{\text{SR}} = U_0 I$. Considering the BS impact on $\sigma_{\delta}$, $\varepsilon_{x,y}$, and $\sigma_{z}$ through the analytical formulas previously mentioned, the other performance parameters, including $\xi_{x,y}$, $\sigma_{W}$, and luminosity per IP ($\mathcal{L}$), were calculated analytically with or without the BS impact. Constrained by the $h$ and a minimum bunch spacing of \SI{25}{\nano\second}, the number of bunches per beam ($n_{b}$) was maximized to 12000 to achieve a minimum $N_{b}$, thereby minimizing the BS impact on the efficiency of monochromatization. The results of the analytical global performance calculation for the monochromatization IR optics based on the “FCC-ee GHC V22 Z” optics are summarized in the second and third columns of Table~\ref{tab1}. Parameters due only to SR are marked by “SR,” while those including the impact of BS are marked by “BS.” The optics labeled “MonochroM ZH4IP” integrates the nonzero $D_{x}^{\ast}$ monochromatization IR optics at all four IPs, while “MonochroM ZH2IP” does so at only two IPs. For comparison, the first column, labeled “Standard ZES,” presents an energy-scaled (ES) optics configuration without realistic implementation. Its performance parameters were calculated after increasing the $E_0$ of the “FCC-ee GHC V22 Z” optics from \SI{45.6}{\giga\electronvolt} to \SI{62.5}{\giga\electronvolt}.

Due to the impact of BS, the $\sigma_{\delta}$ of the “Standard ZES” optics rises from 0.054\% to 0.078\%, resulting in a corresponding blow-up in the $\sigma_{W}$ from \SI{47.47}{\mega\electronvolt} to \SI{68.78}{\mega\electronvolt}. This rise in $\sigma_{\delta}$ also causes the $\sigma_{z}$ to grow from \SI{4.09}{\milli\meter} to \SI{5.82}{\milli\meter}, which increases the $\varphi$ and consequently reduces the $\mathcal{L}$ from \SI{3.54E35}{\per\centi\meter\squared\per\second} to \SI{2.51E35}{\per\centi\meter\squared\per\second}. With nonzero $D_{x}^{\ast}$, the $\varepsilon_{x}$ of the “MonochroM ZH4IP” optics significantly raises from \SI{2.09}{\nano\meter} to \SI{7.78}{\nano\meter} due to the impact of BS, as per Eq.~\ref{eq6}, which decreases the $\lambda$, as per Eq.~\ref{eq3}. However, the $\sigma_{\delta}$ (Eq.~\ref{eq5}) with the BS impact remains nearly the same as the $\sigma_{\delta}$ due only to SR, around 0.055\%, resulting in a reduced $\sigma_{W}$ of $\sim$\SI{50}{\mega\electronvolt} without the monochromatization effect. This reduction, combined with the inverse correlation between horizontal spatial position and energy deviation in the monochromatization scheme, further reduces the $\sigma_{W}$ to \SI{26.42}{\mega\electronvolt}. The decrease in $\mathcal{L}$, compared to the “Standard ZES” optics, is attributed to the increased $\sigma_{x}^{\ast}$, which results from both the increased $\varepsilon_{x}$ and the nonzero $D_{x}^{\ast}$. Nonetheless, the larger $\sigma_{x}^{\ast}$ reduces the $\varphi$, which helps to mitigate the luminosity loss, resulting in a $\mathcal{L}$ of \SI{1.64E35}{\per\centi\meter\squared\per\second}. The “MonochroM ZH2IP” optics, with nonzero $D_{x}$ at only two IPs, exhibits a smaller $\varepsilon_{x}$ but a larger $\sigma_{\delta}$ with the BS impact, compared to the “MonochroM ZH4IP” optics. This results in improved $\sigma_{W}$ and $\mathcal{L}$, measuring \SI{20.19}{\mega\electronvolt} and \SI{1.66E35}{\per\centi\meter\squared\per\second}, respectively.

The same calculations and analyses were also performed for the monochromatization IR optics based on the “FCC-ee GHC V22 $\mathrm{t\bar{t}}$” optics and are summarized in the first three columns of Table~\ref{tab2}. The similar configurations are labeled “Standard TES,” “MonochroM TH4IP,” and “MonochroM TH2IP,” respectively. Compared to the monochromatization IR optics with nonzero $D_{x}^{\ast}$ based on the “FCC-ee GHC V22 Z” optics, those based on the “FCC-ee GHC V22 $\mathrm{t\bar{t}}$” optics exhibit better $\sigma_{W}$ and $\mathcal{L}$ due to the inherently lower $\varepsilon_{x}$ of the baseline standard optics.

\section{FCC-ee GHC monochromatization IR optics design with vertical IP dispersion}
\label{sec4}

\subsection{Monochromatization IR optics design}
\label{subsec4.1}
\begin{figure}[h!]
\centering
\includegraphics[width=0.5\textwidth]{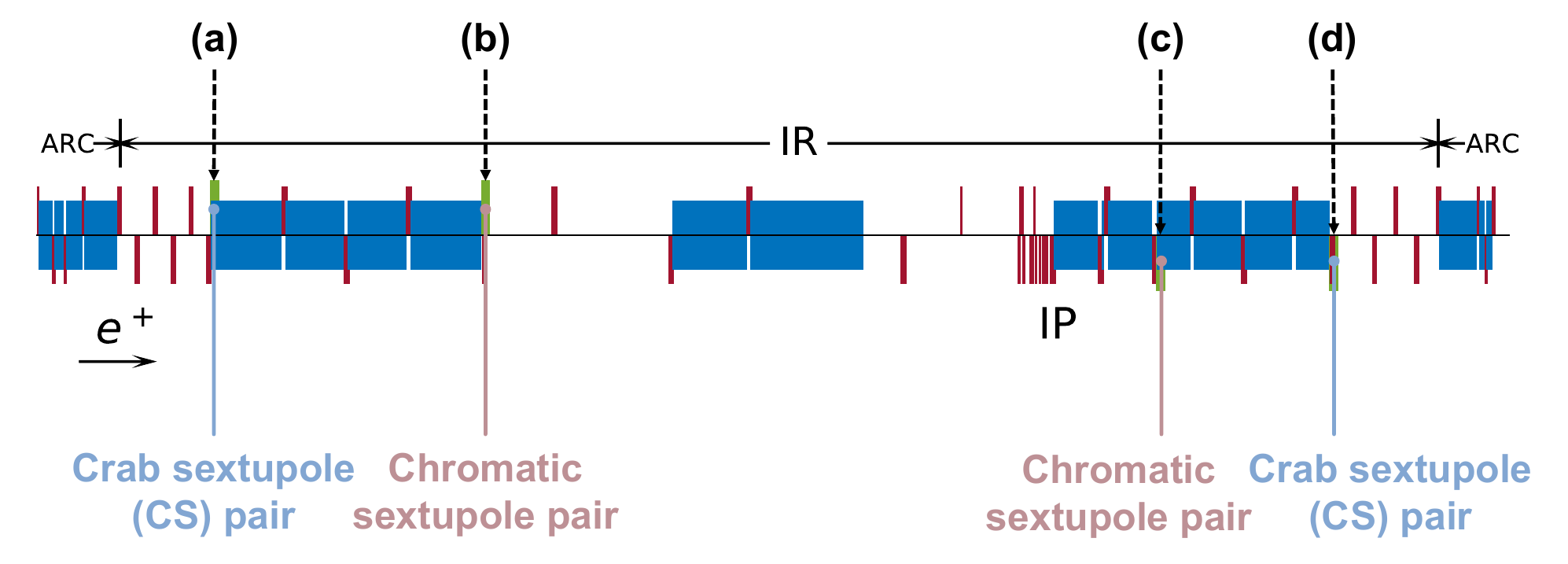}
\caption{Placement of skew quadrupole to generate nonzero vertical IP dispersion. Dipoles, quadrupoles, and sextupoles are shown in blue, red, and green, respectively, while focusing and defocusing elements are positioned above and below the orbit.}\label{fig8}
\end{figure}

The FCC-ee, owing to its extremely low $\varepsilon_{y}$ of \SI{1}{\pico\meter}, features a $\sigma_{y\beta}^{\ast}$ significantly smaller than the $\sigma_{x\beta}^{\ast}$. Therefore, achieving a sufficient $\lambda$, as indicated by Eq.~\ref{eq1}, requires an approximately 100 times smaller $D_{y}^{\ast}$ of $\sim$\SI{1}{\milli\meter}. Inspired by the detector solenoid compensation scheme~\cite{Ramondi:lccsol} for the FCC-ee LCC optics~\cite{Ramondi:lcc2} (an alternative lattice design for the FCC-ee), the introduction of the required $D_{y}^{\ast}$ can be achieved simply without altering the IR orbit by adjusting the strengths of IR skew quadrupoles, which are already necessary for correcting the skew-quadrupole field caused by quadrupole roll errors~\cite{Charles:2023adm}. The locations of these skew quadrupoles are illustrated in Fig.~\ref{fig8}. Two skew quadrupoles, labeled (a) and (d), are installed between the crab sextupole pairs, while the other two, labeled (b) and (c), are placed between chromatic sextupole pairs. The two skew quadrupoles on the same side of the IP, such as (a) and (b), have identical magnitudes but opposite signs in their strengths~\cite{Zhang:2024cepceu,Zhang:2024uxh,Fausgolfe:2024ipac,Zhang:2024phd}.

\begin{figure}[h!]
\centering
  \begin{subfigure}[b]{0.5\textwidth}
     \includegraphics[width=\textwidth]{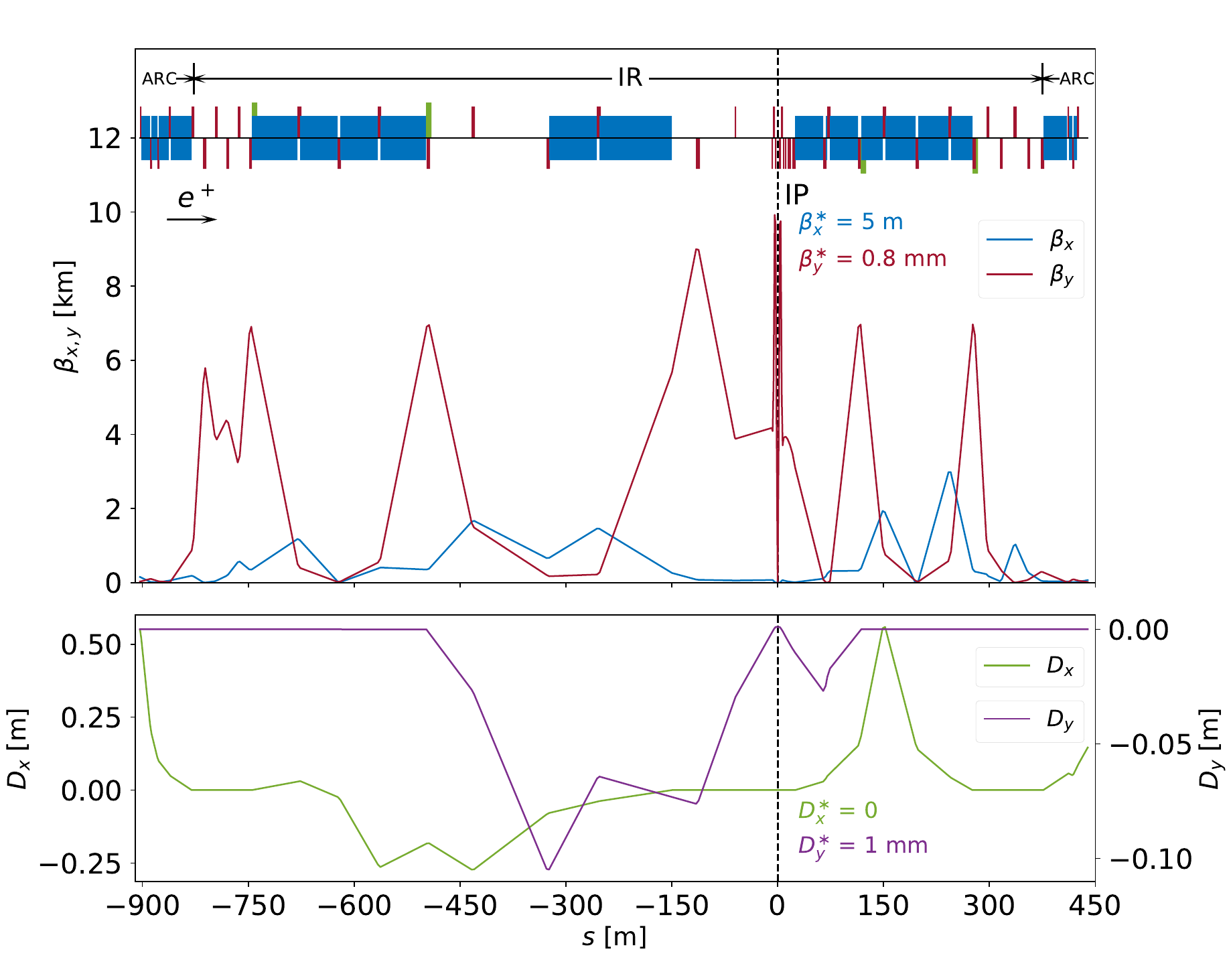}
  \end{subfigure}
  \begin{subfigure}[b]{0.5\textwidth}
     \includegraphics[width=\textwidth]{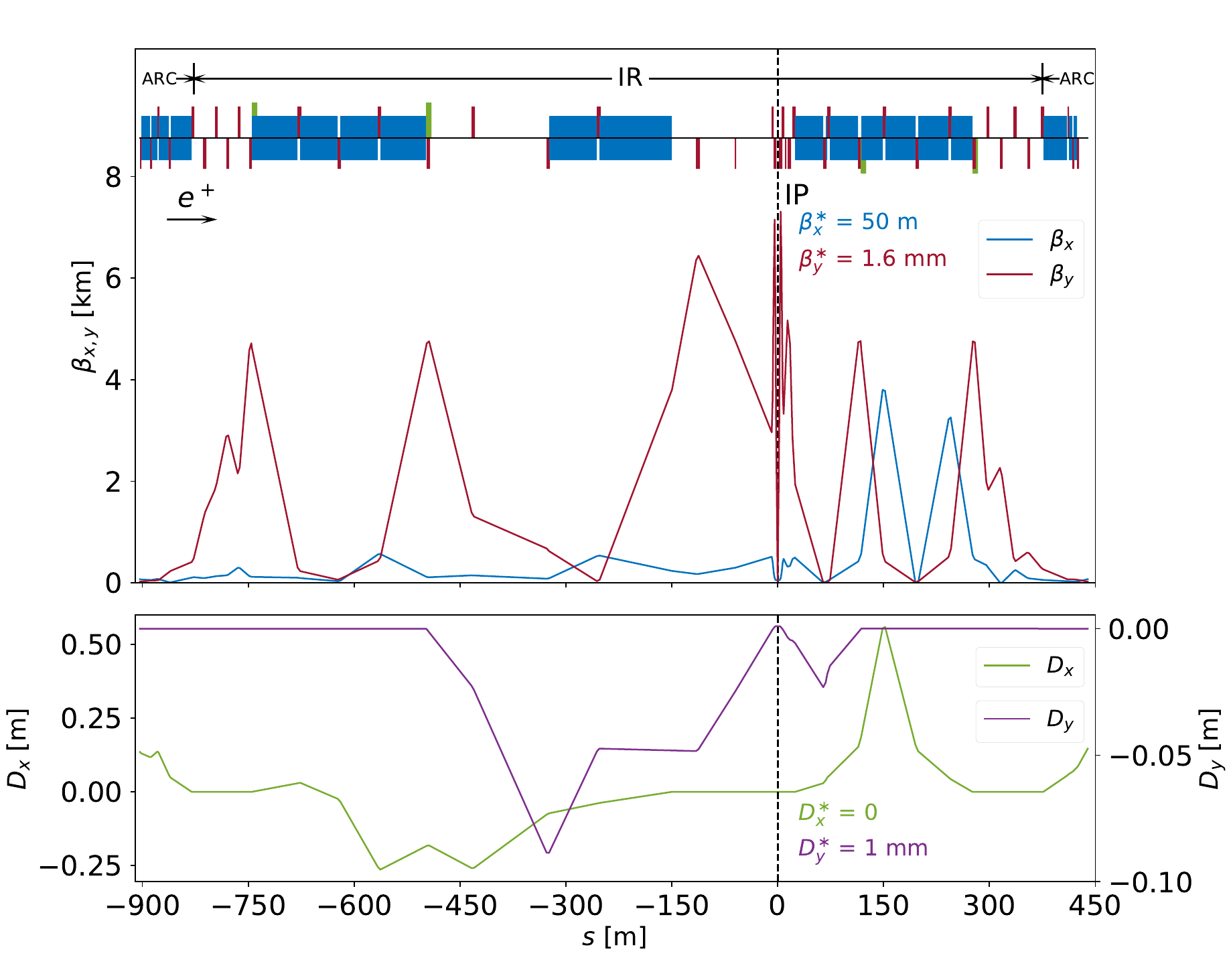}
  \end{subfigure}
\caption{Monochromatization IR lattices and optics with $D_{y}^{\ast}$ of \SI{1}{\milli\meter} based on the “FCC-ee GHC V22 Z” (top) and “FCC-ee GHC V22 $\mathrm{t\bar{t}}$” (bottom) optics calculated using MAD-X. The beam direction is from left to right; the dashed line at $s=0$ marks the IP. In the lattice, dipoles, quadrupoles, and sextupoles are shown in blue, red, and green, respectively, while focusing and defocusing elements are positioned above and below the orbit. In the optics, $\beta_{x,y}$ are shown in blue and red, respectively, while $D_{x,y}$ are shown in green and purple, respectively.}\label{fig9}
\end{figure}

Using this scheme, the monochromatization IR optics designs were developed based on the two types of “FCC-ee GHC V22” optics. While maintaining zero $D_{y}$ at both the entrance and exit of the IR, we matched the $D_{y}^{\ast}$ to \SI{1}{\milli\meter} by varying the strengths of the installed skew quadrupoles. The resulting monochromatization IR lattice and optics based on the “FCC-ee GHC V22 Z” optics are depicted at the top of Fig.~\ref{fig9}. To increase the $\sigma_{x}^{\ast}$, thus mitigating the $\varepsilon_{y}$ blow-up as per Eq.~\ref{eq7}, the $\beta_{x}^{\ast}$ was matched to 50 times the values of the original standard optics by varying the strengths of the final focus quadruples while ensuring all the necessary constraints mentioned in the previous section were satisfied. A similar approach was also applied to the “FCC-ee GHC V22 $\mathrm{t\bar{t}}$” optics. The $\beta_{x}^{\ast}$ was matched to \SI{50}{\meter}, as depicted at the bottom of Fig.~\ref{fig9}.

\subsection{Global optical performance calculation}
\label{subsec4.2}
The global implementation of the monochromatization IR optics with nonzero $D_{y}^{\ast}$ was simpler than the one with nonzero $D_{x}^{\ast}$. It only required the installation of four skew quadrupole magnets between the sextupole pairs in each of the four IRs, with their strengths matched to generate the desired $D_{y}^{\ast}$. Subsequently, the strengths of all quadrupoles in the IR were adjusted to the matched values to increase the $\beta_{x}^{\ast}$. The global optical performances of the monochromatization IR optics were calculated after applying all corrections and SR-RF strategy compensation, as described in the previous section, and are summarized in the second-to-last column of Table~\ref{tab1} and Table~\ref{tab2}, under the label “MonochroM ZV” and “MonochroM TV.”

The $\sigma_{W}$ of the “MonochroM ZV” optics initially decreases to \SI{4.07}{\mega\electronvolt} when the impact of BS is excluded. However, due to BS, the presence of nonzero $D_{y}^{\ast}$ caused a $\varepsilon_{y}$ blow-up from \SI{2.67}{\pico\meter} to \SI{47.3}{\pico\meter}, as per Eq.~\ref{eq7}. This blow-up leads to a reduction of the $\lambda$, as per Eq.~\ref{eq3}. Meanwhile, the $\sigma_{\delta}$ (Eq.~\ref{eq5}) with the BS impact, remains close to the SR-only value of around 0.055\%, owing to the increased $\beta_{x}^{\ast}$, from \SI{0.1}{\meter} to \SI{5}{\meter}. This reduction, along with the monochromatization effect, results in a final $\sigma_{W}$ of \SI{16.22}{\mega\electronvolt}. The decreased $\mathcal{L}$, \SI{1.69E34}{\per\centi\meter\squared\per\second}, is attributed to the increased $\sigma_{x,y}^{\ast}$. Compared to the “MonochroM ZV” optics, the “MonochroM TV” optics features a better $\sigma_{W}$ at the cost of a larger luminosity loss.

\section{FCC-ee GHC monochromatization IR optics design with combined IP dispersion}
\label{sec5}
Considering the potential benefits of combining the two monochromatization schemes discussed above, a monochromatization IR with combination of $D_{x}^{\ast}$ and $D_{y}^{\ast}$ was also developed, based on both types of “FCC-ee GHC V22” optics, by incorporating skew quadrupoles into the monochromatization IR with nonzero $D_{x}^{\ast}$~\cite{Zhang:2024cepceu,Zhang:2024uxh,Fausgolfe:2024ipac,Zhang:2024phd}.
\begin{figure}[h!]
\centering
  \begin{subfigure}[b]{0.5\textwidth}
     \includegraphics[width=\textwidth]{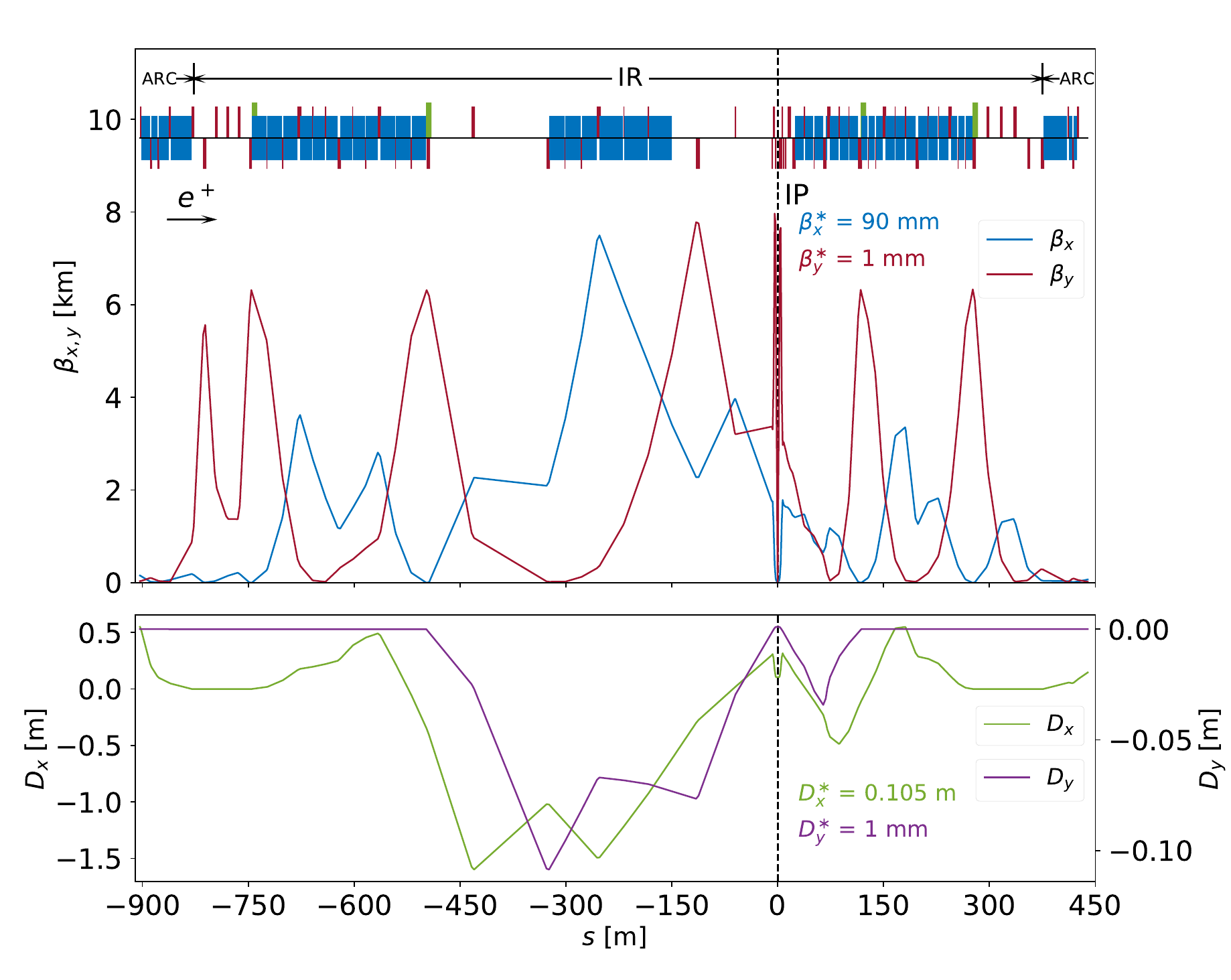}
  \end{subfigure}
  \begin{subfigure}[b]{0.5\textwidth}
     \includegraphics[width=\textwidth]{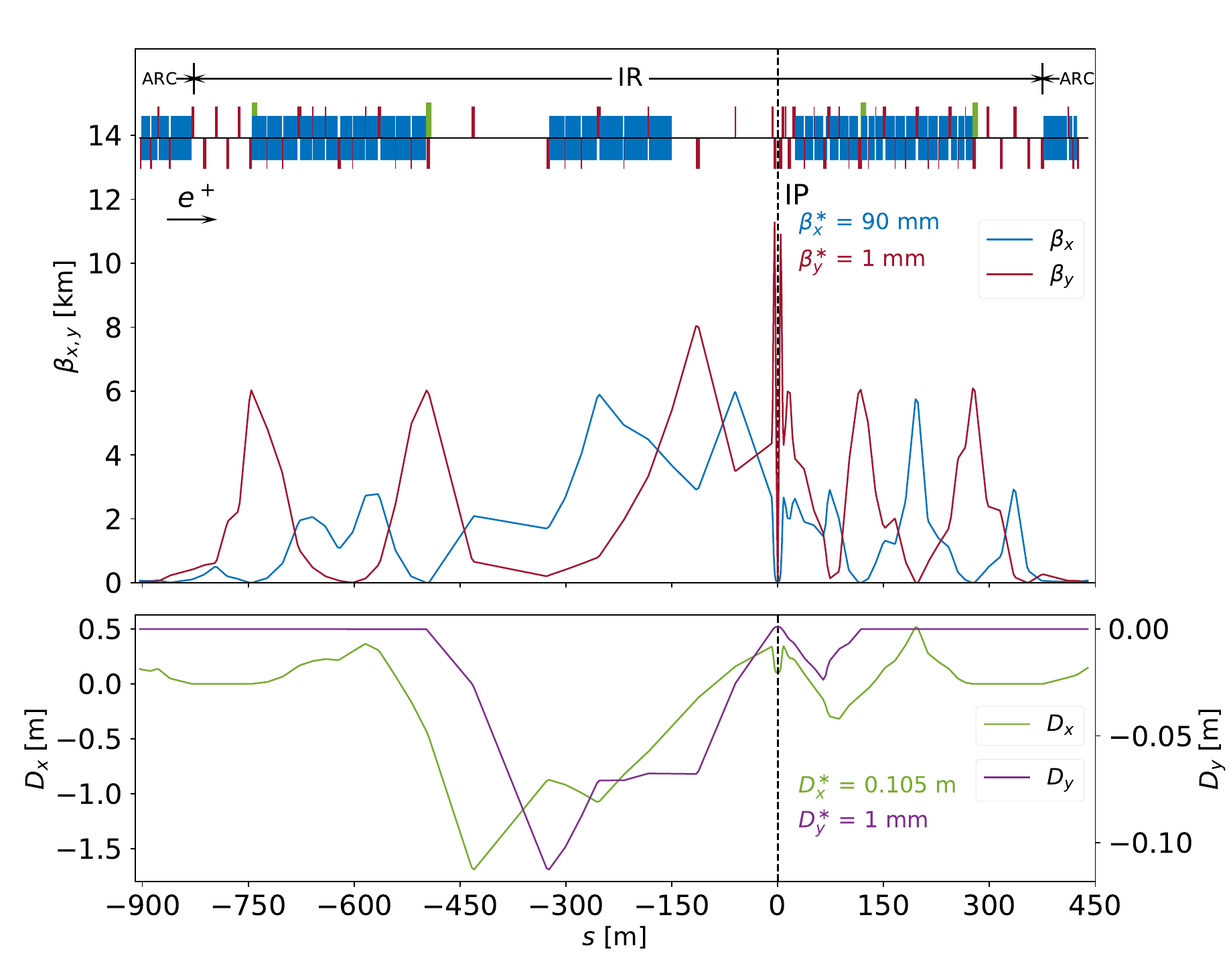}
  \end{subfigure}
\caption{Monochromatization IR lattices and optics with combined $D_{x,y}^{\ast}$ ($D_{x}^{\ast} = \SI{0.105}{\meter}$, $D_{y}^{\ast} = \SI{1}{\milli\meter}$) based on the “FCC-ee GHC V22 Z” (top) and “FCC-ee GHC V22 $\mathrm{t\bar{t}}$” (bottom) optics calculated using MAD-X. The beam direction is from left to right; the dashed line at $s=0$ marks the IP. In the lattice, dipoles, quadrupoles, and sextupoles are shown in blue, red, and green, respectively, while focusing and defocusing elements are positioned above and below the orbit. In the optics, $\beta_{x,y}$ are shown in blue and red, respectively, while $D_{x,y}$ are shown in green and purple, respectively.}\label{fig10}
\end{figure}
The nonzero $D_{y}^{\ast}$, being independent of the $D_{x}$ in linear optics, will be corrected by these skew quadrupoles installed between the sextupole pairs. The IR optics design results are displayed in Fig.~\ref{fig10}, with the corresponding global optical performances summarized in the last column of Table~\ref{tab1} and Table~\ref{tab2}, labeled “MonochroM ZHV” and “MonochroM THV.” Compared to the “MonochroM ZH4IP” and “MonochroM TH4IP” optics, the “MonochroM ZHV” and “MonochroM THV” optics achieve reduced $\sigma_{W}$ of \SI{15.87}{\mega\electronvolt} and \SI{15.69}{\mega\electronvolt}, respectively, thanks to the inclusion of the nonzero $D_{y}^{\ast}$. However, the $\mathcal{L}$ decreases to \SI{1.73E34}{\per\centi\meter\squared\per\second} and \SI{1.72E34}{\per\centi\meter\squared\per\second}, respectively, as a result of the $\varepsilon_{y}$ blow-up caused by the nonzero $D_{y}^{\ast}$.

\begin{table*}[h!]
\centering
\small
\begin{tabular}{>{\raggedright\arraybackslash}p{3.75cm} >{\raggedleft\arraybackslash}p{1cm} >{\centering\arraybackslash}p{1.75cm} >{\centering\arraybackslash}p{1.75cm} >{\centering\arraybackslash}p{1.75cm} >{\centering\arraybackslash}p{1.75cm} >{\centering\arraybackslash}p{1.75cm}}
    \hline
    Parameter & [Unit] & Standard ZES & MonochroM ZH4IP & MonochroM ZH2IP & MonochroM ZV & MonochroM ZHV \\
    \hline
    \# of IPs $n_{\text{IP}}$ & & \multicolumn{5}{c}{4} \\
    Circumference $C$ & [\si{\kilo\meter}] & \multicolumn{5}{c}{91} \\
    Full crossing angle $\theta_{c}$ & [\si{\milli\radian}] & \multicolumn{5}{c}{30} \\
    Beam energy $E_{0}$ & [\si{\giga\electronvolt}] & \multicolumn{5}{c}{62.5} \\
    Energy loss/turn $U_{0}$ & [\si{\mega\electronvolt}] & 137.9 & 142.7 & 140.2 & 137.8 & 142.7 \\
    SR power/beam $P_{\text{SR}}$ & [\si{\mega\watt}] & 50 & 50 & 49 & 50 & 50 \\
    Beam current $I$ & [\si{\milli\ampere}] & 360 & 350 & 350 & 360 & 350 \\
    Bunches/beam $n_{b}$ & & \multicolumn{5}{c}{12000} \\
    Bunch population $N_{b}$ & [$10^{10}$] & 5.70 & 5.54 & 5.54 & 5.70 & 5.54 \\
    Horizontal emittance (SR/BS) $\varepsilon_{x}$ & [\si{\nano\meter}] & 1.33/1.33 & 2.09/7.78 & 1.71/3.56 & 1.34/1.34 & 2.03/7.72 \\
    Vertical emittance (SR/BS) $\varepsilon_{y}$ & [\si{\pico\meter}] & 2.65/2.65 & 4.17/4.17 & 3.42/3.42 & 2.67/47.3 & 4.06/50.5 \\
    Arc cell & & \multicolumn{5}{c}{Long 90/90} \\
    Momentum compaction factor $\alpha_{\text{C}}$ & [$10^{-6}$] & 28.0 & 27.4 & 27.7 & 27.9 & 27.4 \\
    $\beta_{x/y}^{\ast}$ & [\si{\milli\meter}] & 100/0.8 & 90/1 & 90/1 & 5000/0.8 & 90/1 \\
    $D_{x/y}^{\ast}$ & [\si{\meter}] & 0/0 & 0.105/0 & 0.105/0 &  0/0.001 & 0.105/0.001 \\
    Relative energy spread (SR/BS) $\sigma_{\delta}$ & [\%] & 0.054/0.078 & 0.055/0.057 & 0.054/0.064 & 0.054/0.055 & 0.055/0.057 \\
    Bunch length (SR/BS) $\sigma_{z}$ & [\si{\milli\meter}] & 4.09/5.82 & 4.15/4.23 & 4.13/4.74 & 4.10/4.14 & 4.15/4.23 \\
    Piwinski angle (SR/BS) $\varphi$ & [rad] & 5.33/7.58 & 1.05/0.97 & 1.06/1.03 & 0.75/0.76 & 1.05/0.97 \\
    Overlap collision (SR/BS) $2\sigma_{x}^{\ast}/\left(\theta_{c}\beta_{y}^{\ast}\right)$ & & 0.96/0.96 & 3.95/4.36 & 3.89/4.60 & 6.81/6.81 & 3.94/4.35 \\
    RF voltage (400/800 \si{\mega\hertz}) $V_{\text{RF}}$ & [\si{\giga\volt}] &  \multicolumn{5}{c}{0.360/0} \\
    Synchrotron tune $Q_{s}$ & & 0.054 & 0.053 & 0.054 & 0.054 & 0.053 \\
    Long. damping time $\tau_{E,\mathrm{SR}}/T_{\text{rev}}$ & [turns] & 453 & 438 & 446 & 454 & 438 \\
    Horizontal beam-beam (SR/BS) $\xi_{x}$ & & 0.0054/0.0027 & 0.0025/0.0022 & 0.0025/0.0019 & 0.0996/0.0987 & 0.0024/0.0021\\
    Vertical beam-beam (SR/BS) $\xi_{y}$ & & 0.0581/0.0412 & 0.0366/0.0345 & 0.0408/0.0350 & 0.0030/0.0028 & 0.0043/0.0036 \\
    CM energy spread (SR/BS) $\sigma_{W}$ & [\si{\mega\electronvolt}] & 47.47/68.78 & 15.83/26.42 & 14.52/20.19 & 4.07/16.22 & 5.30/15.87\\
    Luminosity/IP \newline (SR/BS) $\mathcal{L}$ & [$10^{34}$ \si{\per\centi\meter\squared\per\second}] & 35.4/25.1 & 17.3/16.4 & 19.3/16.6 & 1.85/1.69 & 2.03/1.73 \\
    \hline
\end{tabular}
\caption{Global optical performance parameters of the monochromatization IR optics based on the “FCC-ee GHC V22 Z” optics.}\label{tab1}
\end{table*}

\clearpage

\begin{table*}[h!]
\centering
\small
\begin{tabular}{>{\raggedright\arraybackslash}p{3.75cm} >{\raggedleft\arraybackslash}p{1cm} >{\centering\arraybackslash}p{1.75cm} >{\centering\arraybackslash}p{1.75cm} >{\centering\arraybackslash}p{1.75cm} >{\centering\arraybackslash}p{1.75cm} >{\centering\arraybackslash}p{1.75cm}}
    \hline
    Parameter &  [Unit]  & Standard TES & MonochroM TH4IP & MonochroM TH2IP & MonochroM TV & MonochroM THV \\
    \hline
    \# of IPs $n_{\text{IP}}$ & & \multicolumn{5}{c}{4} \\
    Circumference $C$ & [\si{\kilo\meter}] & \multicolumn{5}{c}{91} \\
    Full crossing angle $\theta_{c}$ & [\si{\milli\radian}] & \multicolumn{5}{c}{30} \\
    Beam energy $E_{0}$ & [\si{\giga\electronvolt}] & \multicolumn{5}{c}{62.5} \\
    Energy loss/turn $U_{0}$ & [\si{\mega\electronvolt}] & 137.6 & 143.5 & 140.5 & 137.6 & 143.5 \\
    SR power/beam $P_{\text{SR}}$ & [\si{\mega\watt}] & 50 & 50 & 49 & 50 & 50 \\
    Beam current $I$ & [\si{\milli\ampere}] & 360 & 350 & 350 & 360 & 350 \\
    Bunches/beam $n_{b}$ & & \multicolumn{5}{c}{12000} \\
    Bunch population $N_{b}$ & [$10^{10}$] & 5.70 & 5.54 & 5.54 & 5.70 & 5.54 \\
    Horizontal emittance (SR/BS) $\varepsilon_{x}$ & [\si{\nano\meter}] & 0.17/0.17 & 1.48/7.27 & 0.84/4.23 & 0.19/0.19 & 1.48/7.26 \\
    Vertical emittance (SR/BS) $\varepsilon_{y}$ & [\si{\pico\meter}] & 0.35/0.35 & 2.96/2.96 & 1.68/1.68 & 0.37/18.8 & 2.96/50.2 \\
    Arc cell & & \multicolumn{5}{c}{90/90} \\
    Momentum compaction factor $\alpha_{\text{C}}$ & [$10^{-6}$] & 7.31 & 6.92 & 7.12 & 7.31 & 6.92 \\
    $\beta_{x/y}^{\ast}$ & [\si{\milli\meter}] & 1000/1.6 & 90/1 & 90/1 & 50000/1.6 & 90/1 \\
    $D_{x/y}^{\ast}$ & [\si{\meter}] & 0/0 & 0.105/0 & 0.105/0 &  0/0.001 & 0.105/0.001 \\
    Relative energy spread (SR/BS) $\sigma_{\delta}$ & [\%] & 0.054/0.076 & 0.055/0.057 & 0.054/0.057 & 0.054/0.055 & 0.055/0.057 \\
    Bunch length (SR/BS) $\sigma_{z}$ & [\si{\milli\meter}] & 3.86/5.49 & 4.05/4.20 & 3.95/4.12 & 3.86/3.95 & 4.05/4.20 \\
    Piwinski angle (SR/BS) $\varphi$ & [rad] & 4.39/6.25 & 1.03/0.96 & 1.02/0.98 & 0.60/0.61 & 1.03/0.96 \\
    Overlap collision (SR/BS) $2\sigma_{x}^{\ast}/\left(\theta_{c}\beta_{y}^{\ast}\right)$ & & 0.55/0.55 & 3.94/4.36 & 3.86/4.19 & 4.03/4.03 & 3.94/4.36 \\
    RF voltage (400/800 \si{\mega\hertz}) $V_{\text{RF}}$ & [\si{\giga\volt}] &  \multicolumn{5}{c}{0.170/0} \\
    Synchrotron tune $Q_{s}$ & & 0.015 & 0.014 & 0.014 & 0.015 & 0.014 \\
    Long. damping time $\tau_{E,\mathrm{SR}}/T_{\text{rev}}$ & [turns] & 454 & 436 & 445 & 454 & 436 \\
    Horizontal beam-beam (SR/BS) $\xi_{x}$ & & 0.0593/0.0300 & 0.0025/0.0022 & 0.0027/0.0024 & 0.8192/0.8087 & 0.0025/0.0022\\
    Vertical beam-beam (SR/BS) $\xi_{y}$ & & 0.2389/0.1701 & 0.0439/0.0410 & 0.0597/0.0561 & 0.0055/0.0051 & 0.0043/0.0036\\
    CM energy spread (SR/BS) $\sigma_{W}$ & [\si{\mega\electronvolt}] & 47.45/67.58 & 13.41/25.75 & 10.25/20.95 & 2.16/14.63 & 4.52/15.69\\
    Luminosity/IP \newline (SR/BS) $\mathcal{L}$ & [$10^{34}$ \si{\per\centi\meter\squared\per\second}] & 72.8/51.9 & 20.9/19.5 & 28.3/26.6 & 1.68/1.56 & 2.05/1.72 \\
    \hline
\end{tabular}
\caption{Global optical performance parameters of the monochromatization IR optics based on the “FCC-ee GHC V22 $\mathrm{t\bar{t}}$” optics.}\label{tab2}
\end{table*}

With the global optical performance of all proposed FCC-ee GHC monochromatization IR optics summarized in Table~\ref{tab1} and Table~\ref{tab2}, we observe that the $\varphi$ in all the monochromatization schemes is reduced to $\sim$1 due to the increased $\sigma_{x}^{\ast}$. In the baseline collision scheme of the FCC-ee, a large $\varphi$ is necessary to mitigate the hourglass effect~\cite{Shatilov:2018drj}. To estimate the hourglass reduction factor, the ratio of the collision overlap area $2\sigma_{x}^{\ast}/\theta_{c}$ and $\beta_{y}^{\ast}$, expressed as $2\sigma_{x}^{\ast}/\left( \theta_{c} \beta_{y}^{\ast} \right)$, is included as it characterizes the hourglass effect~\cite{Bogomyagkov:2017tpk}. For all the monochromatization cases, this ratio is $\sim$4, corresponding to a $R_{\mathrm{hg}}$ of $\sim$0.2 from Ref.~\cite{Bogomyagkov:2017tpk}. This issue could potentially be mitigated by increasing $\beta_{y}^{\ast}$, which will be addressed in future studies through a comprehensive optimization that considers all relevant effects. Additionally, it is noteworthy that the theoretical values of $\xi_{y}$ for all these optics meet the constraint $\xi_{y} \leq 0.14$~\cite{oide:ghc2022} demonstrating the feasibility of implementing monochromatization under the crossing-angle collision configuration at the FCC-ee. Further beam-beam simulations, including the impact of the $D_{x,y}^{\ast}$, are currently under investigation and will be integrated into the Xsuite code~\cite{xsuite}. Future in-deep studies will focus on a step-by-step dynamic aperture optimization. This optimization will involve fine-tuning the arc sextupole families based on particle tracking results to ensure reliable performance for these monochromatization optics.

\section{FCC-ee GHC monochromatization physics performance evaluations}
\label{sec6}
Two critical optical parameters, $\sigma_{W}$ and $\mathcal{L}_{\text{int}}$, serve as key indicators of the physics performance in collider experiments. To accurately assess the performance of the FCC-ee GHC monochromatization IR optics, we first calculated their $\sigma_{W}$ and $\mathcal{L}$ using the simulation tool Guinea-Pig~\cite{Schulte:1997nga}, incorporating a preliminary estimation of the beam-beam effects. In this calculation, the particle distribution at the IP was not a realistic distribution generated by accelerator optics, but modeled as an ideal Gaussian distribution, comprising 40000 particles and defined by the following global optical performance parameters: $E_0$, $\sigma_{\delta}$, $\varepsilon_{x,y}$, $\beta_{x,y}^{\ast}$, $D_{x,y}^{\ast}$, $\sigma_{z}$, and $\theta_{c}$. Since Guinea-Pig simulates collisions per bunch crossing, it omits $n_{b}$ or $f_{r}$ when calculating $\mathcal{L}$. Therefore, when simulating the collision scenario at the IP for each optics design using Guinea-Pig, $N_{b}$ needs to be adjusted to an appropriate value so that the $\mathcal{L}$ calculated by Guinea-Pig under the assumed head-on collision configuration matches the analytical values of the corresponding optics design. Then, the $\mathcal{L}$ under the crossing-angle collision configuration should be calculated using the adjusted $N_{b}$. Besides, Guinea-Pig is unable to simulate BS effects in the presence of non-zero $D_{x,y}^{\ast}$. Therefore, during the simulations, the initial particle distributions were generated using theoretical values of $\sigma_{\delta}$, $\varepsilon_{x,y}$, and $\sigma_{z}$ that already account for the impact of BS. To avoid double-counting the BS effects, the corresponding simulation functionality was disabled in Guinea-Pig.

\begin{figure}[h!]
\centering
\includegraphics[width=0.5\textwidth]{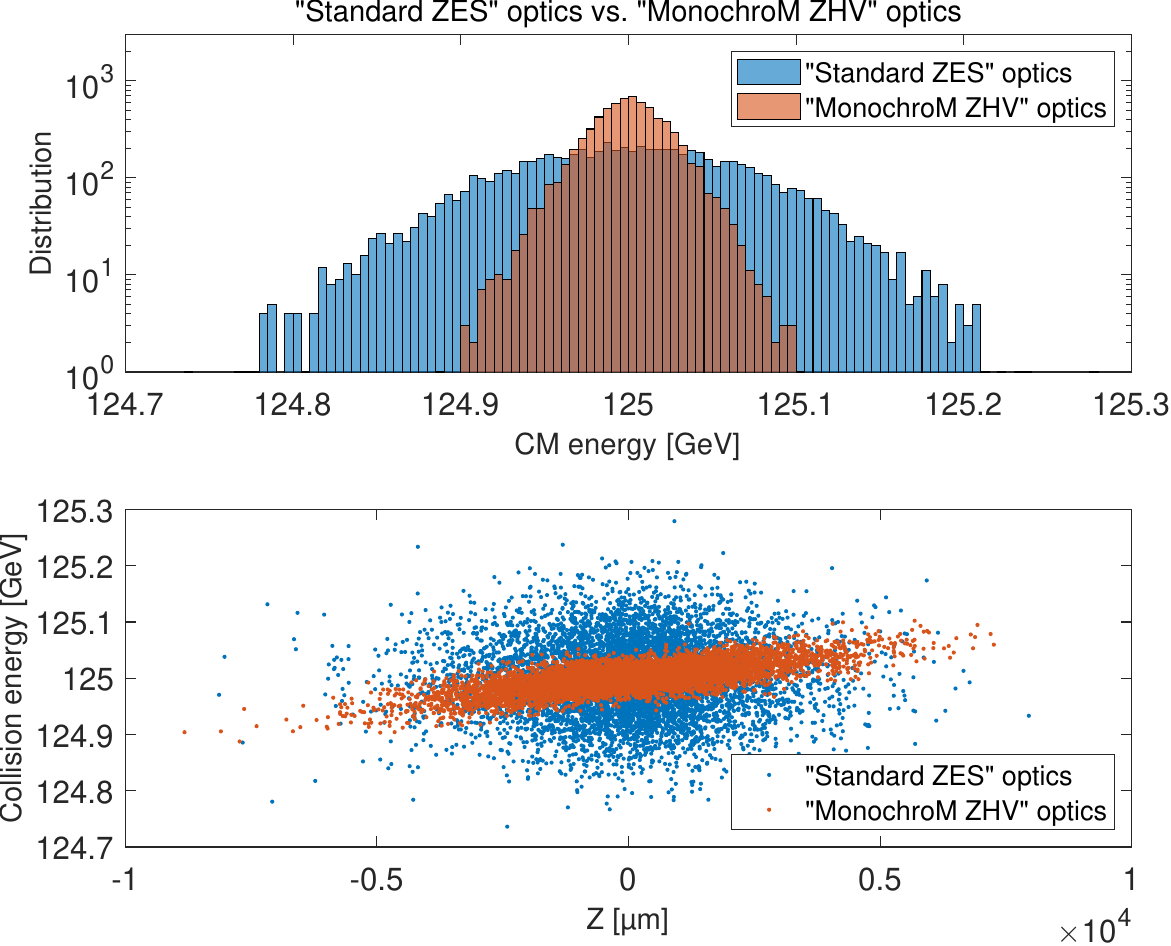}
\caption{Comparison of the CM energy distribution (top) and its correlation with the longitudinal collision position (bottom) between the “Standard ZES” (blue) and “MonochroM ZHV” (orange) optics, as simulated using Guinea-Pig.}\label{fig11}
\end{figure}

Guinea-Pig simulates and records all the colliding $e^+ e^-$ pairs along with the following parameters of each pair: energy of the $e^+$, energy of the $e^-$, horizontal collision position, vertical collision position, longitudinal collision position, horizontal angle of the $e^+$, vertical angle of the $e^+$, horizontal angle of the $e^-$, and vertical angle of the $e^-$. The sum of the $e^+$ and $e^-$ energy gives the CM energy of each colliding pair. By analyzing the distribution of the CM energy, we could calculate the $\sigma_{W}$. The top of Fig.~\ref{fig11} compares the CM energy distribution of the “Standard ZES” and “MonochroM ZHV” optics. It is evident that the $\sigma_{W}$ is reduced with opposite correlation between transverse spatial position and energy deviation, thanks to the nonzero $D_{x,y}^{\ast}$. The correlation between collision energy and longitudinal position, depicted at the bottom of Fig.~\ref{fig11}, demonstrates that the monochromatization efficiency is diminished due to the $\theta_{c}$, in accordance with Eq.~\ref{eq3} and Eq.~\ref{eq4}. The $\sigma_{W}$ and $\mathcal{L}$ of all monochromatization IR optics are summarized in two tables: Table~\ref{tab3} for those based on the “FCC-ee GHC V22 Z” optics and Table~\ref{tab4} for those based on the “FCC-ee GHC V22 $\mathrm{t\bar{t}}$” optics.

\begin{table*}[h!]
\centering
\small
\begin{tabular}{>{\raggedright\arraybackslash}p{3.75cm} >{\raggedleft\arraybackslash}p{1cm} >{\centering\arraybackslash}p{1.75cm} >{\centering\arraybackslash}p{1.75cm} >{\centering\arraybackslash}p{1.75cm} >{\centering\arraybackslash}p{1.75cm} >{\centering\arraybackslash}p{1.75cm}}
    \hline
    Parameter &  [Unit]  & Standard ZES & MonochroM ZH4IP & MonochroM ZH2IP & MonochroM ZV & MonochroM ZHV \\
    \hline
    CM energy spread $\sigma_{W}$ & [\si{\mega\electronvolt}] & 69.15 & 27.29 & 24.12 & 25.82 & 21.52 \\
    Luminosity/IP $\mathcal{L}$ & [$10^{34}$ \si{\per\centi\meter\squared\per\second}] & 44.6 & 15.0 & 18.4 & 1.46 & 1.42 \\
    Integrated luminosity/ \newline IP/year $\mathcal{L}_{\text{int}}$ & [\si{\ab}] & 5.35 & 1.80 & 2.21 & 0.18 & 0.17 \\
    \hline
\end{tabular}
\caption{$\sigma_{W}$, $\mathcal{L}$, and $\mathcal{L}_{\text{int}}$ of the monochromatization IR optics based on the “FCC-ee GHC V22 Z” optics.}\label{tab3}
\end{table*}

\begin{table*}[h!]
\centering
\small
\begin{tabular}{>{\raggedright\arraybackslash}p{3.75cm} >{\raggedleft\arraybackslash}p{1cm} >{\centering\arraybackslash}p{1.75cm} >{\centering\arraybackslash}p{1.75cm} >{\centering\arraybackslash}p{1.75cm} >{\centering\arraybackslash}p{1.75cm} >{\centering\arraybackslash}p{1.75cm}}
    \hline
    Parameter &  [Unit]  & Standard TES & MonochroM TH4IP & MonochroM TH2IP & MonochroM TV & MonochroM THV \\
    \hline
    CM energy spread $\sigma_{W}$ & [\si{\mega\electronvolt}] & 67.51 & 26.32 & 23.68 & 19.52 & 20.44 \\
    Luminosity/IP $\mathcal{L}$ & [$10^{34}$ \si{\per\centi\meter\squared\per\second}] & 71.2 & 17.8 & 24.5 & 1.37 & 1.42 \\
    Integrated luminosity/ \newline IP/year $\mathcal{L}_{\text{int}}$ & [\si{\ab}] & 8.54 & 2.14 & 2.94 & 0.16 & 0.17 \\
    \hline
\end{tabular}
\caption{$\sigma_{W}$, $\mathcal{L}$, and $\mathcal{L}_{\text{int}}$ of the monochromatization IR optics based on the “FCC-ee GHC V22 $\mathrm{t\bar{t}}$” optics.}\label{tab4}
\end{table*}

Regarding the $\mathcal{L}_{\text{int}}$, based on the assumptions laid out in the FCC-ee CDR~\cite{Abada2019FCCeeTL}, which include 185 physics days per year and a physics efficiency of 75\%, a $\mathcal{L}$ of \SI{E35}{\per\centi\meter\squared\per\second} yields \SI{1.2}{\ab} per year per IP. Using this relation, the $\mathcal{L}_{\text{int}}$ values were calculated and are listed in the corresponding tables.

\begin{figure}[h!]
\centering
\includegraphics[width=0.5\textwidth]{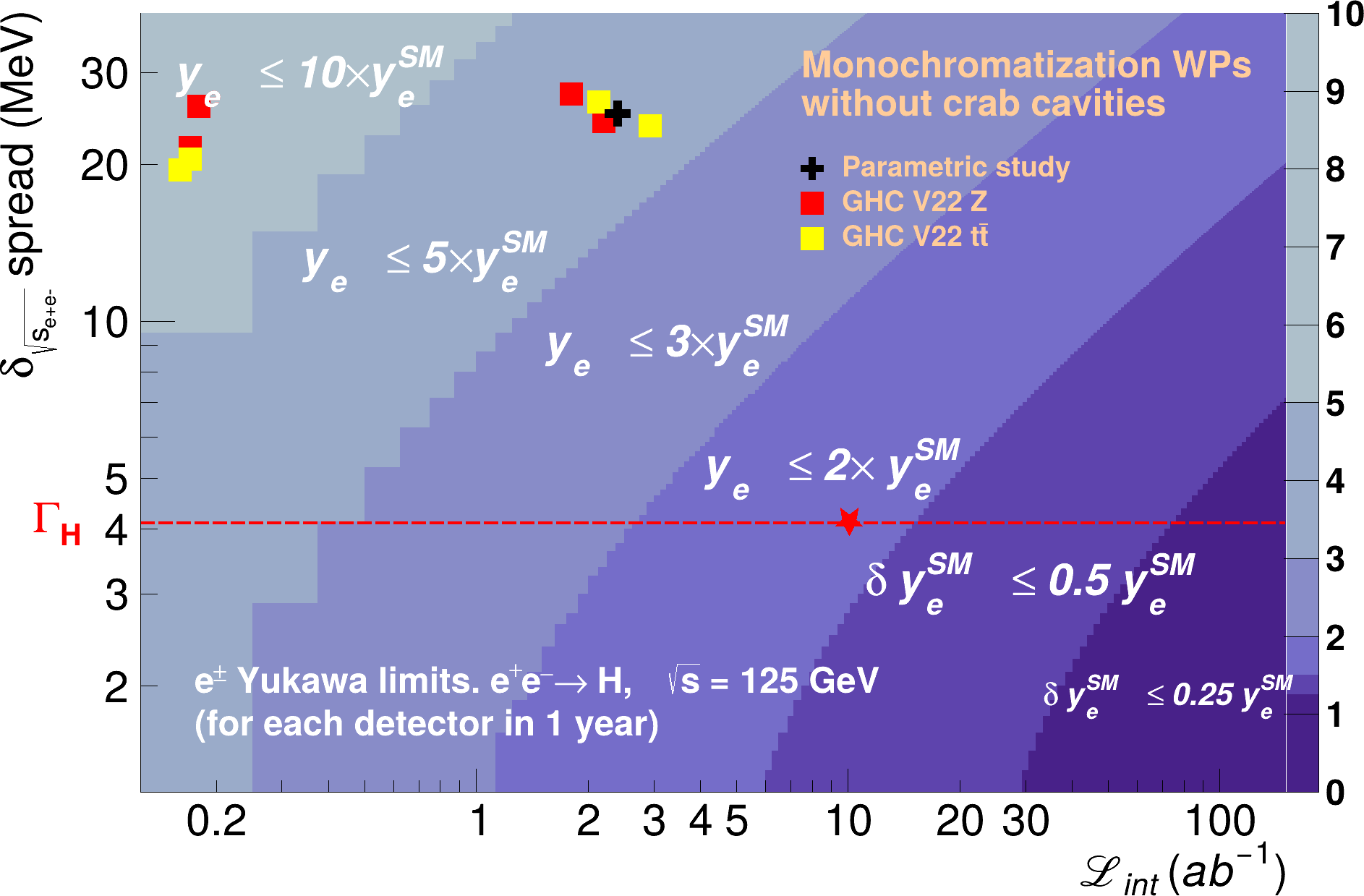}
\caption{Upper limits contours (95\% CL) on the electron Yukawa $y_{e}$ coupling in the $\delta_{\sqrt{s}}$ vs. $\mathcal{L}_{\text{int}}$ plane, per FCC-ee IP and per year. The red and yellow points indicate the monochromatization points listed in Tables~\ref{tab3} and Table~\ref{tab4}, respectively. The red star over the $\delta_{\sqrt{s}} = \SI{4.1}{\mega\electronvolt}$ dashed line indicates the baseline point assumed in the original physics simulation analysis~\cite{dEnterria:2021xij}. The black cross indicates the previously achieved working point with the parametrized studies~\cite{ValdiviaGarcia:2022nks,Faus-Golfe:2021udx}.}\label{fig12}
\end{figure}

For physics performance evaluation of the FCC-ee monochromatization mode, large samples of simulated signal ($e^+ e^- \rightarrow \text{H} \rightarrow \text{XX}$) and associated background ($e^+ e^- \rightarrow \text{Z}^{\ast} \rightarrow \text{XX}$) events have been generated with the \textsc{Pythia} 8 Monte Carlo code~\cite{dEnterria:2021xij}. This was done for 11 Higgs decay channels, employing a multivariate analysis that utilizes Boosted Decision Trees to discriminate between signal and background events. This analysis leads to the identification of two promising final states in terms of statistical significance (in units of standard deviations $\sigma$): $\text{H} \rightarrow gg$ and $\text{H} \rightarrow \text{WW}^{\ast} \rightarrow \ell \nu 2j$. Taking a monochromatization benchmark that provides an ideal CM energy spread $\delta_{\sqrt{s}}$ of \SI{4.1}{\mega\electronvolt} and a $\mathcal{L}_{\text{int}}$ of \SI{10}{\ab}, an upper limit on the electron Yukawa coupling was achieved at 1.6 times of the SM Higgs \textit{s}-channel cross section: $\lvert y_{e} \rvert < 1.6\,\lvert y_{e}^{SM} \rvert$ at a 95\% confidence level (CL) per IP per year, represented by the red star in Fig.~\ref{fig12}. The black cross represents the $\delta_{\sqrt{s}}$-$\mathcal{L}_{\text{int}}$ benchmark of the FCC-ee monochromatization self-consistent parameters, corresponding to a $\delta_{\sqrt{s}}$ of \SI{25}{\mega\electronvolt} and a $\mathcal{L}_{\text{int}}$ of \SI{2.76}{\ab}.

The physics performances of the proposed FCC-ee GHC monochromatization IR optics were evaluated by establishing their $\delta_{\sqrt{s}}$-$\mathcal{L}_{\text{int}}$ benchmarks on the 95\% CL upper limits contours for $y_{e}$ coupling, as illustrated in Fig.~\ref{fig12}. These benchmarks were determined by the $\sigma_{W}$ and $\mathcal{L}_{\text{int}}$ of each type of monochromatization settings, as listed in Table~\ref{tab3} and Table~\ref{tab4}. The red benchmarks refer to those based on the “FCC-ee GHC V22 Z” optics, while the yellow ones are based on the “FCC-ee GHC V22 $\mathrm{t\bar{t}}$” optics. The physics performances of all designed monochromatization IR optics with nonzero $D_{x}^{\ast}$ are comparable to or even exceed that of the FCC-ee self-consistent parameters (black cross). The “MonochroM TH2IP” optics achieves the best $\delta_{\sqrt{s}}$-$\mathcal{L}_{\text{int}}$ benchmark, with $\sigma_{W} = \SI{23.68}{\mega\electronvolt}$ and $\mathcal{L}_{\text{int}} = \SI{2.94}{\ab}$. This corresponds to an upper limit (95\% CL) of $\lvert y_{e} \rvert < 3.2\,\lvert y_{e}^{SM} \rvert$ for the Higgs-electron coupling, per IP per year. For the monochromatization IR optics with nonzero $D_{y}^{\ast}$, the best physics performance benchmark is achieved with the “MonochroM TV” optics, yielding an upper limit (95\% CL) of $\lvert y_{e} \rvert < 7.1\,\lvert y_{e}^{SM} \rvert$. The “MonochroM ZHV” and “MonochroM THV” optics, which incorporate combined IP dispersion, do not provide improved physics performance compared to the optics with only nonzero $D_{x}^{\ast}$. This is primarily due to the increased luminosity loss caused by the $\varepsilon_{y}$ blow-up when accounting for the impact of BS.

\begin{figure}[h!]
\centering
\includegraphics[width=0.5\textwidth]{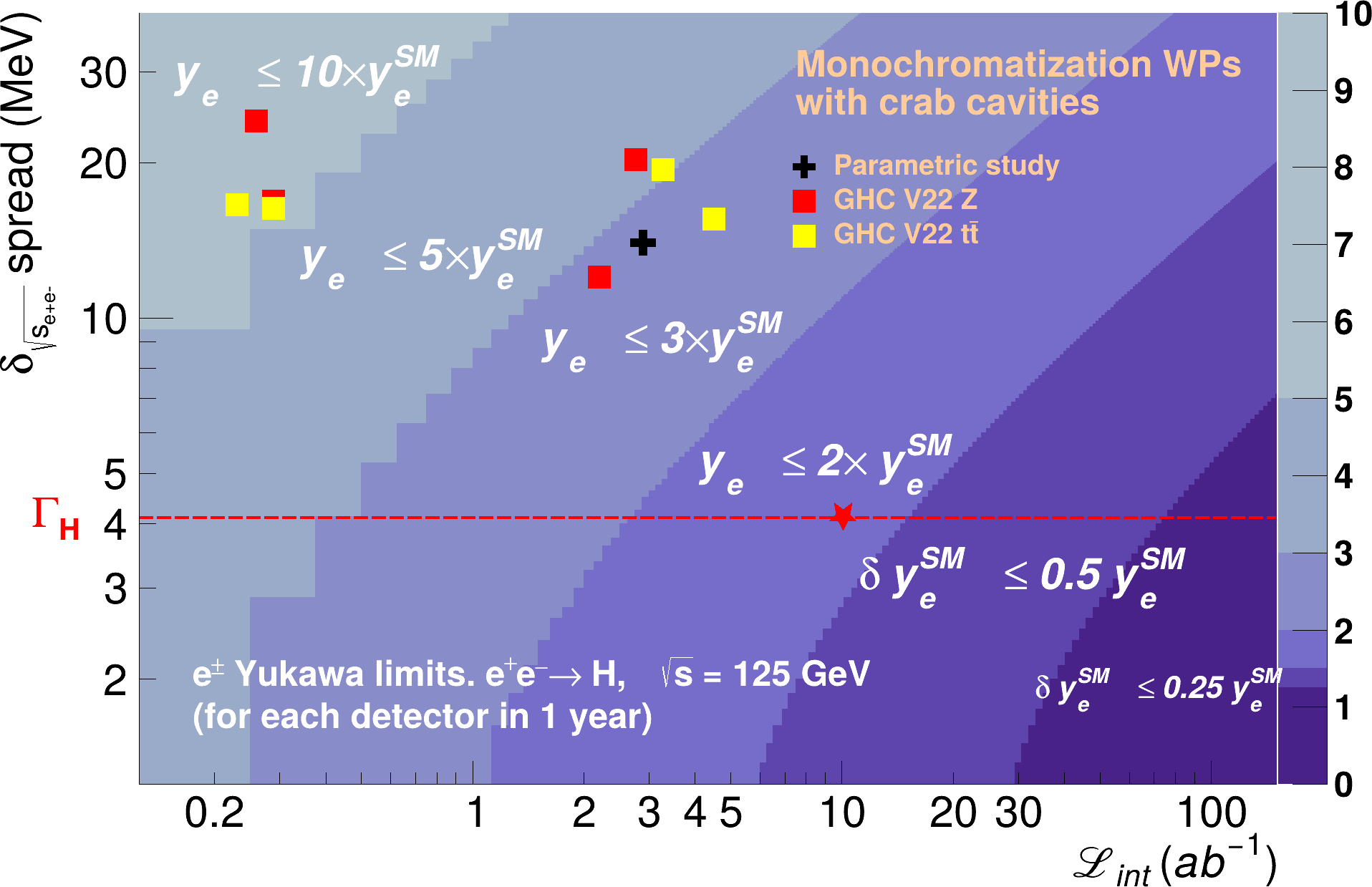}
\caption{Upper limits contours (95\% CL) on the electron Yukawa $y_{e}$ coupling in the $\delta_{\sqrt{s}}$ vs. $\mathcal{L}_{\text{int}}$ plane, per FCC-ee IP and per year. The red and yellow points indicate the monochromatization points with assumed crab cavities based on the “FCC-ee GHC V22 Z” and “FCC-ee GHC V22 $\mathrm{t\bar{t}}$” optics, respectively~\cite{Zhang:2024phd}. The red star over the $\delta_{\sqrt{s}} = \SI{4.1}{\mega\electronvolt}$ dashed line indicates the baseline point assumed in the original physics simulation analysis~\cite{dEnterria:2021xij}. The black cross indicates the previously achieved working point with the parametrized studies~\cite{ValdiviaGarcia:2022nks,Faus-Golfe:2021udx}.}\label{fig13}
\end{figure}

With the same analysis under the head-on collision configuration including crab cavities, the physics performances of all proposed monochromatization IR optics are further improved~\cite{Zhang:2024phd}, as shown in Fig.~\ref{fig13}. The best $\delta_{\sqrt{s}}$-$\mathcal{L}_{\text{int}}$, achieved with the “MonochroM TH2IP” optics, yielding $\sigma_{W} = \SI{15.54}{\mega\electronvolt}$ and $\mathcal{L}_{\text{int}} = \SI{4.51}{\ab}$, indicates an upper limit of $\lvert y_{e} \rvert < 2.7\,\lvert y_{e}^{SM} \rvert$. It should be noted that the use of crab cavities is not part of the baseline collision scheme for FCC-ee, as the design assumes a large $\varphi$, which does not need or even contradicts crab cavities.

\section{Conclusions and perspectives}
\label{sec7}
Monochromatization is a straightforward conceptual idea, but its practical implementation in a collider is challenging, especially when it is not integrated into the initial IR optics design as a dedicated operational mode. Research on monochromatization remains preliminary and has not yet reached the stage for full implementation and experimental validation. To address this, a flexible lattice design that accommodates both operational modes—with and without monochromatization—is essential. This paper demonstrates that monochromatization is a promising approach for significantly improving CM energy resolution in future high-energy circular $e^+ e^-$ colliders, particularly for precision Higgs boson measurements at the FCC-ee. By exploring various monochromatization schemes and optimizing the IR optics for each scenario, we have laid the groundwork for future experimental implementation and validation. The results indicate that generating either nonzero $D_{x}^{\ast}$, $D_{y}^{\ast}$, or a combination of $D_{x}^{\ast}$/$D_{y}^{\ast}$ are all viable approaches, offering the potential to significantly reduce the $\sigma_{W}$ and enhance the physics reach of the FCC-ee. Although the scheme with nonzero $D_{x}^{\ast}$ provides the best physics performance compared to the others, its complex optics design, which involves splitting all the IR horizontal dipoles into three sets with thin quadrupoles inserted between for compatibility with the standard operation mode, requires further simplification. While the physics performance of the nonzero-$D_{y}^{\ast}$ scheme is less favorable, it is easier to implement without altering the IR orbit, making it feasible for existing low-energy $e^+ e^-$ colliders. Without the $\varepsilon_{y}$ blow-up due to BS, this scheme could potentially achieve better performance in such settings. The hybrid scheme achieves a lower $\sigma_{W}$, but its physics performance does not improve as expected due to the increased luminosity loss. Additionally, its optics design may encounter potential challenges related to betatron coupling.

Several avenues for future research have been identified through this paper. First, the FCC-ee GHC monochromatization IR optics will be upgraded to the latest version of the continuously evolving FCC-ee GHC optics, with the Version 2024 being the most up-to-date. The optical parameters for the monochromatization mode will continue to be optimized for enhanced performance. Second, the dynamic aperture optimization for these new types of monochromatization optics will be carried out step-by-step by adjusting arc sextupole families according to particle tracking results. This will allow for beam-beam simulations with nonzero IP dispersion in the code Xsuite~\cite{xsuite}, incorporating the hourglass and BS effects, rather than relying solely on analytical evaluations. Third, further comparisons and implementations of monochromatization in more symmetric IRs, such as those in the FCC-ee LCC optics and CEPC optics, will be explored. Finally, studies are underway to validate the monochromatization concept experimentally in existing low-energy circular $e^+ e^-$ colliders, such as BEPC II, DA$\Phi$NE, and SuperKEKB. These efforts will be essential to realizing the full potential of monochromatization in future collider projects.

\section*{Acknowledgments}

The author gratefully acknowledges the financial support provided by the China Scholarship Council (CSC) for this work. Special thanks are extended to R. Aleksan for early discussions on the physics motivation behind these studies. The author also wishes to thank A. Blondel, P. Janot, D. Shatilov, J. Keintzel, G. Wilkinson, and J. Wenninger for their valuable discussions within the EPOL FCC-ee working group, and expresses further appreciation to M. Hofer, K. André, and X. Buffat for their insightful private communications and discussions during the FCC-ee accelerator design meetings.




 \bibliographystyle{elsarticle-num-names} 
 \bibliography{bibliography}
 \biboptions{sort&compress}






\end{document}